\documentclass[twocolumn,times]{aastex63}
\hypersetup{linkcolor=red,citecolor=blue,filecolor=cyan,urlcolor=magenta}
\usepackage{rotating}
\usepackage{multirow}
\usepackage{amsmath}

\newcommand{\kms}{\mbox{km\,s$^{-1}$}}
\newcommand{\methanol}{\mbox{$\rm CH_3OH$}}

\newcommand{\hcop}{\mbox{$\rm HCO^+$}}
\newcommand{\sotwo}{\mbox{$\rm SO_2$}}

\newcommand{\thirteenco}{\mbox{$\rm ^{13}CO$}}

\newcommand{\mujypbm}{\mbox{$\mu$Jy\,beam$^{-1}$}}
\newcommand{\jypbm}{\mbox{Jy\,beam$^{-1}$}}
\newcommand{\lsun}{\mbox{\,$L_\odot$}}
\newcommand{\zsun}{\mbox{\,$Z_\odot$}}
\newcommand{\msun}{\mbox{\,$M_\odot$}}

\graphicspath{{./}{figures/}}

\received{}
\revised{}
\accepted{}
\submitjournal{ApJ}

\shorttitle{}
\shortauthors{Cheng et al.}

\begin{document}

\title{A High Resolution View of Three Massive Young Stellar Objects in the Magellanic Clouds}

\correspondingauthor{Yu Cheng}
\email{ycheng.astro@gmail.com}

\author[0000-0002-8691-4588]{Yu Cheng}
\affil{National Astronomical Observatory of Japan, 2-21-1 Osawa, Mitaka, Tokyo, 181-8588, Japan}

\author[0000-0002-6907-0926]{Kei E. I. Tanaka}
\affiliation{Department of Earth and Planetary Sciences, Institute of Science Tokyo, Meguro, Tokyo, 152-8551, Japan}

\author[0000-0002-0095-3624]{Takashi Shimonishi}
\affiliation{Institute of Science and Technology, Niigata University, Ikarashi-ninocho 8050, Nishi-ku, Niigata 950-2181, Japan}

\author[0000-0001-7511-0034]{Yichen Zhang}
\affiliation{Department of Astronomy, School of Physics and Astronomy, Shanghai Jiao Tong University, 800 Dongchuan Road, Shanghai 200240, People's Republic of China}
\affiliation{State Key Laboratory of Dark Matter Physics, School of Physics and Astronomy, Shanghai Jiao Tong University, Shanghai 200240, People's Republic of China}
\affiliation{Key Laboratory for Particle Astrophysics and Cosmology (MOE)/Shanghai Key Laboratory for Particle Physics and Cosmology, Shanghai 200240, People's Republic of China}

\author[0000-0002-8389-6695]{Suinan Zhang}
\affiliation{Department of Earth and Planetary Sciences, Institute of Science Tokyo, Meguro, Tokyo, 152-8551, Japan}
\affiliation{National Astronomical Observatory of Japan, 2-21-1 Osawa, Mitaka, Tokyo, 181-8588, Japan}

\author[0000-0002-2026-8157]{Kenji Furuya}
\affiliation{RIKEN Pioneering Research Institute, 2-1 Hirosawa, Wako-shi, Saitama 351-0198, Japan}

\author[0000-0001-5817-6250]{Asako Sato}
\affiliation{Institut de Ciències de l'Espai (ICE-CSIC), Campus UAB, Can Magrans S/N, E-08193 Cerdanyola del Vallès, Catalonia, Spain}

\begin{abstract}

The Magellanic Clouds provide an excellent laboratory to study massive star formation beyond typical Galactic environments. Using ALMA band 7 long-baseline observations, we present the highest resolution (approximately 0\farcs{03}, $1500{\rm\:au}$) view to date of three massive young stellar objects (MYSOs) in the Magellanic Clouds, i.e., Lh03 and Lh09 in the LMC, and S07 in the SMC. The 0.87~mm continuum, with a sensitivity of $\sim30\:\mu$\jypbm, reveals varying levels of fragmentation, from a single {compact source} (S07), a few (Lh03), to a cluster of {compact sources} (Lh09). In Lh09, we identified 17 core {candidates} with lowest measured mass of 2.6\msun, and their de-projected separations peak around 6000~au. All three targets exhibit high velocity CO(3--2) outflows, including detection of a collimated bipolar jet driven by a core in Lh09. Spectral lines from HCO$^+$, SO, and SO$_2$ are detected in all sources. We observe velocity gradients that likely trace rotating toroids (Lh03) or more complex motions (Lh09, S07). {Taken together, these phenomena observed on $\sim1500$~au scales are qualitatively similar to those seen in Galactic MYSOs, suggesting a degree of universality in the basic processes of massive star formation across different environments. Nevertheless, the substantial source-to-source diversity in continuum morphology and dense-gas kinematics motivates larger high-resolution surveys to determine how these properties vary with source characteristics and environment.}

\end{abstract}


\section{Introduction}\label{sec:intro}

Across the universe, massive stars ($\gtrsim$10~\msun) play dominant roles in the dynamical and chemical evolution of the interstellar medium through their feedback, synthesis and dispersal of heavy elements. Despite their importance, the large distance and short lived nature makes it difficult to study their birth and evolution, and the mechanisms of their formation remain subjects of intense investigation \citep{Tan14}. Over the past decade, our understanding of high mass star formation has advanced rapidly through the combined progress of numerical simulations, galactic surveys, and increasingly high resolution interferometric observations \citep[e.g.,][]{Rosen20, Molinari25, Olguin26,Yang26}. Growing evidence suggests that high mass stars can form through disk-mediated accretion, broadly analogous to their lower mass counterparts, although the process is more strongly coupled to large scale mass supply \citep[e.g.,][]{Kumar20,Xu24}, and intense feedback at later stages \citep[e.g.,][]{Tanaka18}.

Despite its galaxy scale importance, our current understanding of massive star formation is still built mainly on Milky Way benchmarks, which spans a relatively narrow range of environmental parameter space. Metal-poor conditions, for example, are common in the early Universe, as metallicity increases over cosmic time. A lower metallicity can reshape the early assembly and feedback of massive stars by regulating the dust abundance, opacity, thermal balance, ionization, and magnetic coupling of star forming gas. {Protostellar disk models relevant to massive star-forming conditions predict a non-monotonic dependence of fragmentation on metallicity, with particularly strong fragmentation for $Z \sim 10^{-5}$--$10^{-2}\,Z_\odot$, where reduced dust opacity can bring optical depths in dense disk gas toward unity, allowing efficient radiative cooling \citep{Tanaka14,Matsukoba22}. More general cluster simulations also suggest enhanced small-scale fragmentation, smaller disk sizes, and a higher close-binary fraction toward lower metallicity \citep{Bate19,Elsender21}. The feedback associated with massive protostars may also change. Magnetically driven protostellar outflows are expected over a broad metallicity range as long as the magnetic field remains sufficiently coupled to the gas \citep{Higuchi19}, while photoevaporation becomes increasingly important for massive stars at $Z \lesssim 10^{-2}\,Z_\odot$ because of the reduced absorption of ionizing photons by dust \citep{Tanaka18}. }


The Large and Small Magellanic clouds (LMC \& SMC) are excellent laboratories to expand our understanding of massive star formation beyond the Galaxy. As satellite galaxies of the Milky Way, they are situated at a relatively close distance (49.97$\pm$1.11~kpc, \citealt{Pietrzynski13}; 62.1$\pm$1.9~kpc for SMC, \citealt{Graczyk14}), exhibit a notably low metallicity (1/3--1/2~$Z_\odot$ for LMC, \citealt{Russell92,Rolleston02}; 1/10--1/4~$Z_\odot$ for SMC, \citealt{Russell92,Choudhury18}), and host a large number of active massive star formation sites \citep{Seale09, oliveira2013}. With the capability of the Atacama Large Millimeter/submillimeter Array (ALMA), progress has been made in characterizing the dynamical and chemical properties of high mass star formation regions down to $\lesssim$0.1~pc scale. Studies have revealed phenomena familiar from Galactic massive star formation, including filaments, hot cores and outflows \citep[][]{Tokuda19,Tokuda22,Sewilo22,Shimonishi16,Shimonishi20,Shimonishi23,HamedaniGolshan24,Shimonishi26}. Observations reaching scales of a few 1000~au remain rare \citep{McLeod24,Traficante26}, yet such observations are essential for understanding massive star formation, as they directly probe the disk or inner envelope, and the most recent outflow ejection events.

Here we report the highest resolution observations to date of three massive young stellar objects (MYSOs) in the Magellanic Clouds with ALMA, reaching physical scales of $\sim1500$~au in the LMC and $\sim1800$~au in the SMC. The sample is selected from the ``Magellanic Clouds Outflow and Chemistry Survey (MAGOS)'', which observed 30 sources in the LMC and 10 sources in the SMC (see \citealt{Shimonishi26} for the LMC survey). The three objects, Lh03 and Lh09 in the LMC \citep{Shimonishi26}, and S07 in the SMC \citep{Shimonishi23}, were selected because they have high luminosities ($>{2}\times10^4~\lsun$), pronounced chemical richness on $\sim0.1$~pc scales, and evident outflow activities. The target properties are listed in \autoref{table:target_info}.


\section{Observation}\label{sec:obs}

\begin{deluxetable*}{ccccccccccc}
\tabletypesize{\scriptsize}
\renewcommand{\arraystretch}{1.0}
\tablecaption{Target properties of three MYSOs in Magellanic Clouds \label{table:target_info}}
\tablehead{
\colhead{Source} & \colhead{Other identifiers} & \colhead{Galaxy} & \colhead{R.A.} & \colhead{Decl.} & \colhead{$L_{\rm bol}$}  & \colhead{$v_{\rm sys}^b$} &\colhead{Resolution}  & \colhead{1 $\sigma$ cont. sensitivity} & \colhead{5$\sigma$ mass sensitivity$^{a}$ } & \colhead{5$\sigma$ mass sensitivity$^{a}$ }\\
\colhead{} & \colhead{} & \colhead{} & \colhead{(J2000)} & \colhead{(J2000)} & \colhead{($10^4 L_\odot$)}& \colhead{(\kms)} & \colhead{(\arcsec$\times$\arcsec)} & \colhead{(\mujypbm)}  & \colhead{(\msun @50K)}  & \colhead{(\msun @20K)}
}
\startdata
Lh09 & H72.96-69.39(1) & LMC & 04:51:53.32    & $-$69:23:28.71    &    20.80  &  {234.0} & 0.041$\times$0.030  & 44 & 2.3  & 7.6\\
Lh03 & ST16(2) & LMC & 05:19:12.30    & $-$69:09:07.4      & 11.60   &  {265.0} & 0.038$\times$0.035  & 35  & 1.8 & 6.1\\
S07  & Y246(3), \#18(4) & SMC & 00:54:03.44    & $-$73:19:38.5    & 2.8     &  {163.0} & 0.041$\times$0.030 &  24  &  3.1 & 10.3 \\
\enddata
\tablenotetext{a}{Mass sensitivity assuming the 0.87~mm continuum is dominated by optically thin thermal dust emission. For Lh03 and Lh09 we assume a metallicity of 0.4\zsun, and for S07 we assume 0.2\zsun. See the text for details regarding the assumptions about dust properties.}
\tablenotetext{b}{
The systemic velocities are adopted from previous ALMA observations at $\sim0.1$~pc resolution. For Lh09 and Lh03, the values are determined from the peak velocities of the HCO$^+$(4--3) spectra \citep{Tokuda23}. For S07, the systemic velocity is determined from Gaussian fitting of the SO line \citep{Tokuda22}, consistent with the value reported by \citet{Shimonishi23}.
}
\tablerefs{
(1) \citet{Nayak19};
(2) \citet{Shimonishi20};
(3) \citet{Sewilo13};
(4) \citet{oliveira2013}.
}
\end{deluxetable*}

The observations were conducted with ALMA in Band~7 in Cycle 10 (Project ID 2023.1.01629.S, PI: Cheng Y.) during October 2023. We employed the C43-8 configuration to achieve a high spatial resolution of $\sim$0\farcs{03} ($\sim$1500~au for LMC, 1800~au for SMC). The maximum recoverable scale (MRS) is 0\farcs{41}, or $\sim$0.1~pc at the distance of Magellanic Clouds. We set the central frequency of the correlator sidebands to be 347.06~GHz, 357.44~GHz and 359.04~GHz for SPW1, SPW2 and SPW3; SPW0 is split into two spectral windows centered on 345.65~GHz and 344.66~GHz. This setup covers spectral lines from various molecular lines including CO, SO, \sotwo, \methanol, and SiO.

The raw data were calibrated with the data reduction pipeline using {\it{CASA}} 6.5.4.9. The continuum visibility data were constructed with all line-free channels. We performed imaging with the {\it tclean} task in {\it CASA} using a Briggs weighting scheme with a robust parameter of 0.5. Self-calibration was performed for Lh09, which had the highest signal-to-noise ratio among the targets. We carried out one round of phase only self calibration
on an execution-block timescale, and applied the resulting calibration solutions
to both the continuum and spectral line data. The resulting resolution and sensitivity are listed in \autoref{table:target_info}. {The spectral line cubes have velocity channel widths of approximately 0.82--0.85~km~s$^{-1}$ across the observed spectral windows. The typical rms sensitivity of the spectral line cubes is approximately \(1.0\ {\rm mJy\,beam^{-1}}\) per channel.}

\section{Results}\label{sec:results}
\subsection{Continuum and source identification}\label{sec:cont}

\begin{figure*}[ht!]
\centering
\includegraphics[width=0.9\textwidth]{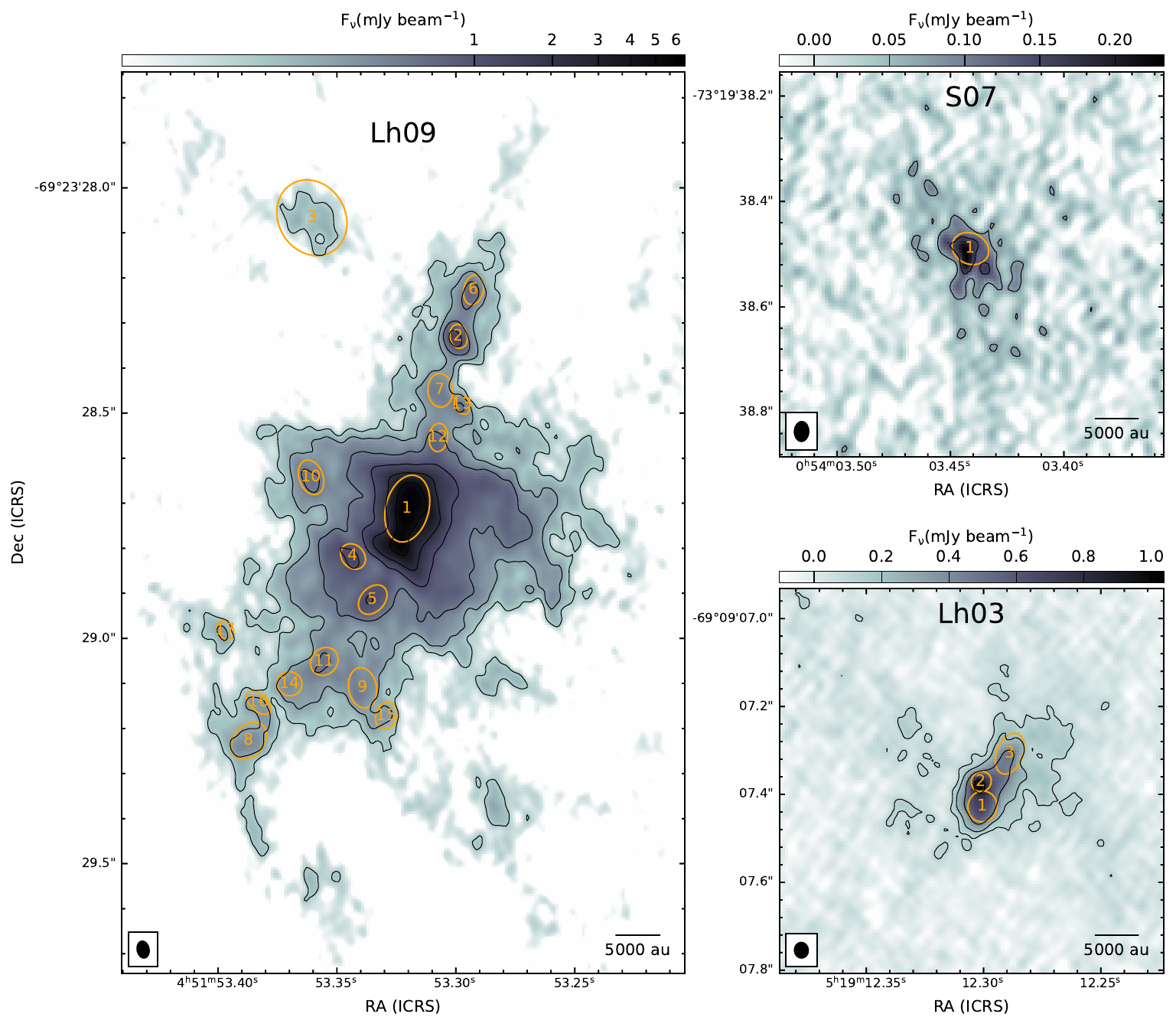}
\caption{ALMA 0.87~mm continuum maps for the three MYSOs: Lh09, Lh03 and S07. All panels are displayed with the same physical scale. The colorscale and contours illustrate the continuum emission with contour levels of $\sigma \times$ (3, 6, 12, 25, 50, 100), where the continuum sensitivity $\sigma$ is listed in \autoref{table:target_info}. Orange ellipses mark the {\it getsf}-identified compact sources, with sizes corresponding to the major and minor FWHM axes. The beam size is indicated in the bottom left corner for each panel. For Lh09, a logarithmic stretch is applied to the colorscale to better display the full dynamic range of the emission.  }
\label{fig:cont}
\end{figure*}

In \autoref{fig:cont}, we present the 0.87~mm continuum images for the three sources. While we have achieved similar flux density sensitivities among the sample, the continuum morphology varies significantly, with Lh09 exhibiting {the richest substructure}, and Lh03 and S07 showing only one or a few {compact sources}. Lh09 is primarily north-south oriented, with an elongation of about 1\farcs{5} (or 0.4~pc). For Lh03 and S07, detections of continuum are mainly limited to the central part within 0\farcs{5}. Lh03 shows a bright peak with extension to the south and northwest; S07 shows a relatively weak detection with a $\sim$9$\sigma$ peak, with some extension in the northeast-southwest direction. 

We use the {\it getsf} method \citep{Men21} to identify {compact sources} from the continuum map. {\it Getsf} is a source extraction method for astronomical images commonly used in Galactic surveys of massive protoclusters \citep[e.g.,][]{Pouteau22,Cheng24}. For this analysis, we adopt a maximum source size of 0.1~pc. The extraction results are shown in \autoref{fig:cont}. {\it Getsf} also provides estimates of source sizes and flux densities, which are listed in \autoref{sec:core_prop}. Source sizes, flux densities, and peak flux densities are measured on the observed images after subtraction of locally interpolated backgrounds and deblending of overlapping sources, with source footprints defining the measurement regions; integrated fluxes are obtained within the footprints, while source sizes are characterized by the major and minor half-maximum axes. {We refer to the compact continuum sources returned by {\it getsf} as core candidates.} The sources are numbered in descending order of their flux densities, e.g., Lh09c1, Lh09c2, and so on. Overall, {\it getsf} identifies 1 core candidate in S07, 3 core candidates in Lh03, and 17 core candidates in Lh09.

If we assume the emission is dominated by optically thin isothermal dust emission, the flux sensitivity can be converted to mass sensitivity using the equation:
\begin{equation}
M_{\rm dust} = \frac{d^2F_{\nu}}{\kappa_{\nu}B_{\nu}(T_{\rm dust})},
\end{equation}
where \(d\) is the distance to the source, \(F_{\nu}\) is the observed flux density, \(B_{\nu}\) is the Planck function, \(T_{\rm dust}\) is the dust temperature, and \(\kappa_{\nu}\) is the dust opacity at the observed frequency. We adopt \(\kappa_{\nu} = 1.84 \, \text{g cm}^{-2}\) from \citet{Ossenkopf94} (thin ice mantles, \(\rm 10^6 \, \text{cm}^{-3}\)). The dust mass is then converted into gas mass assuming gas-to-dust ratios of 250 and 500 for the LMC and SMC, respectively. These values are obtained by scaling the Galactic gas-to-dust ratio of 100 \citep{Bohlin78} with metallicity. For this, we adopt metallicities of 0.4 \(\text{Z}_{\odot}\) for LMC and 0.2 \(\text{Z}_{\odot}\) for the SMC \citep{Russell92}, throughout the paper. {We note, however, that the gas-to-dust ratio does not necessarily scale linearly with metallicity. Observational estimates in the Magellanic Clouds show substantial variations with spatial scale and methodology \citep[e.g.,][]{Welty12,RomanDuval14,Takekoshi18}, and the gas-to-dust ratio appropriate for dense protostellar material remains poorly constrained. We therefore regard the adopted values as fiducial estimates for setting the absolute mass scale. The uncertainty in the gas-to-dust ratio alone introduces a systematic uncertainty of order a factor of two in the derived masses.}

{As listed in \autoref{table:target_info}, our 5$\sigma$ mass sensitivity is approximately 3~\msun\ for sources in SMC and 2~\msun\ for sources in LMC assuming a temperature of 50~K. This limit increases to higher masses ($\sim$10~\msun{} for SMC, and $\sim$7~\msun{} for LMC) if a lower temperature of 20~K is assumed. If these compact structures correspond to individual star-forming cores and a 30\% core-to-star efficiency is adopted \citep[e.g.,][]{Alves07,Tanaka17}, our observations would be sensitive to cores capble of forming high- ($\gtrsim$8~\msun) or intermediate-mass \citep[2--8~\msun,][]{Beltran15} stars, as well as some lower mass cores if they are already sufficiently heated. }

\subsection{Line detections and source characterization}

\begin{figure*}[ht!]
\centering
\includegraphics[width=0.9\textwidth]{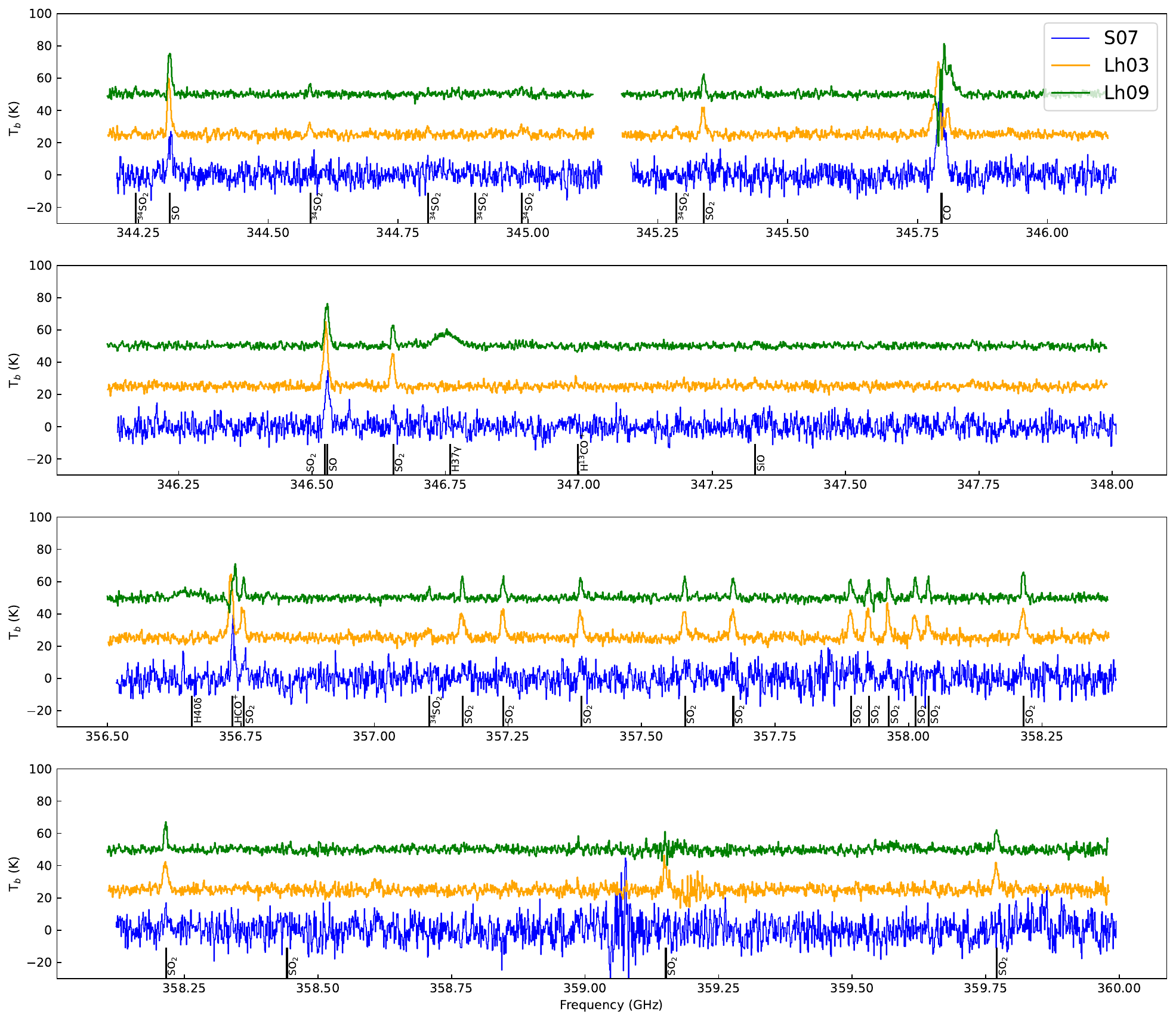}
\caption{ Spectra of the three targets taken at representative positions (the position of core S07c1, Lh03c2, Lh09c1) averaged over a 0\farcs{1} radius (for Lh03c2 and Lh09c1) or 0\farcs{03} radius (for S07c1) aperture. Molecules of the identified lines are labeled with black lines.  
}
\label{fig:spec} 
\end{figure*} 

\autoref{fig:spec} presents representative spectra for the three sources. The spectrum was extracted from the center of each FOV (corresponding to the positions of Lh09c1, Lh03c2, S07c1) and averaged over a 0\farcs{1} ($\sim$5000~au) radius aperture, which encompasses most of the continuum detection for Lh03 and primarily Lh09c1 for Lh09. For S07c1 we adopt a smaller aperture of 0\farcs{03} in radius to avoid dilution since its line emission is very compact. In addition to CO, all three sources are detected in \hcop{}(4-3) at 356.73422~GHz, and SO transition 8$_8$-7$_7$ (344.31061~GHz) and 9$_8$-8$_7$ (346.52848~GHz), and multiple \sotwo{} lines from 345.33 to 359.77~GHz. These \sotwo{} lines have upper energy levels ranging from 48 to 321~K, indicating presence of warm/hot gas. For the two sources in the LMC, a few lines from the rare isotope molecule $\rm ^{34}SO_2$ are also detected in the first spectral window. In addition, in Lh09c1 we have also detected two high order hydrogen recombination lines (HRLs, $\rm H(37)\gamma$ and $\rm H(40)\delta$), suggesting the presence of ionized gas. {The spatial distributions and kinematics of the main molecular tracers are discussed in \autoref{sec:kinematics}.}

{For the compact continuum sources with sufficiently strong \sotwo{} emission, we further use the multiple detected transitions to constrain their gas temperatures. We extract the mean spectrum within the FWHM returned by {\it getsf} for each source and perform forward modeling of the \sotwo{} lines assuming local thermodynamic equilibrium (LTE) and a common filling factor for all transitions (see \autoref{sec:app_temperature} for details). The free parameters include the source velocity $v_{\rm cen}$, velocity dispersion $\sigma_v$, \sotwo{} column density $N$(\sotwo), excitation temperature $T_{\rm ex}$, and filling factor. This approach is similar to a rotation diagram analysis \citep{Goldsmith99}, but accounts for potentially optically thick lines and simultaneously estimates additional parameters. We perform the fit for core candidates with sufficiently high signal-to-noise (S/N) ratios. Specifically, we require the \sotwo{} $13_{2,12}-12_{1,11}$ transition at 345.3385~GHz to have an S/N ratio greater than 4. This enables gas temperature estimation for all 3 core candidates in Lh03, and 5 out of 17 core candidates in Lh09. The measured $T_{\rm ex}$ ranges from 69~K to 96~K, as listed in \autoref{sec:core_prop}.}

{We use the measured temperatures to estimate the core masses assuming optically thin dust emission with assumptions described above. For core candidates without temperature measurements we adopt 50~K. The resulting masses are listed in \autoref{table:core_info}. Note that the mass is likely overestimated given potential free-free contamination in 0.87~mm. Excluding Lh09-c1, which is mostly likely dominated by free-free emission (see \autoref{sec:free-free}), the masses range from 2.6~\msun{} to 18.6~\msun, with a median of 7.8~\msun. }

\subsection{High velocity CO emission and outflows}

\begin{figure*}[ht!]
\centering
\includegraphics[width=1.\textwidth]{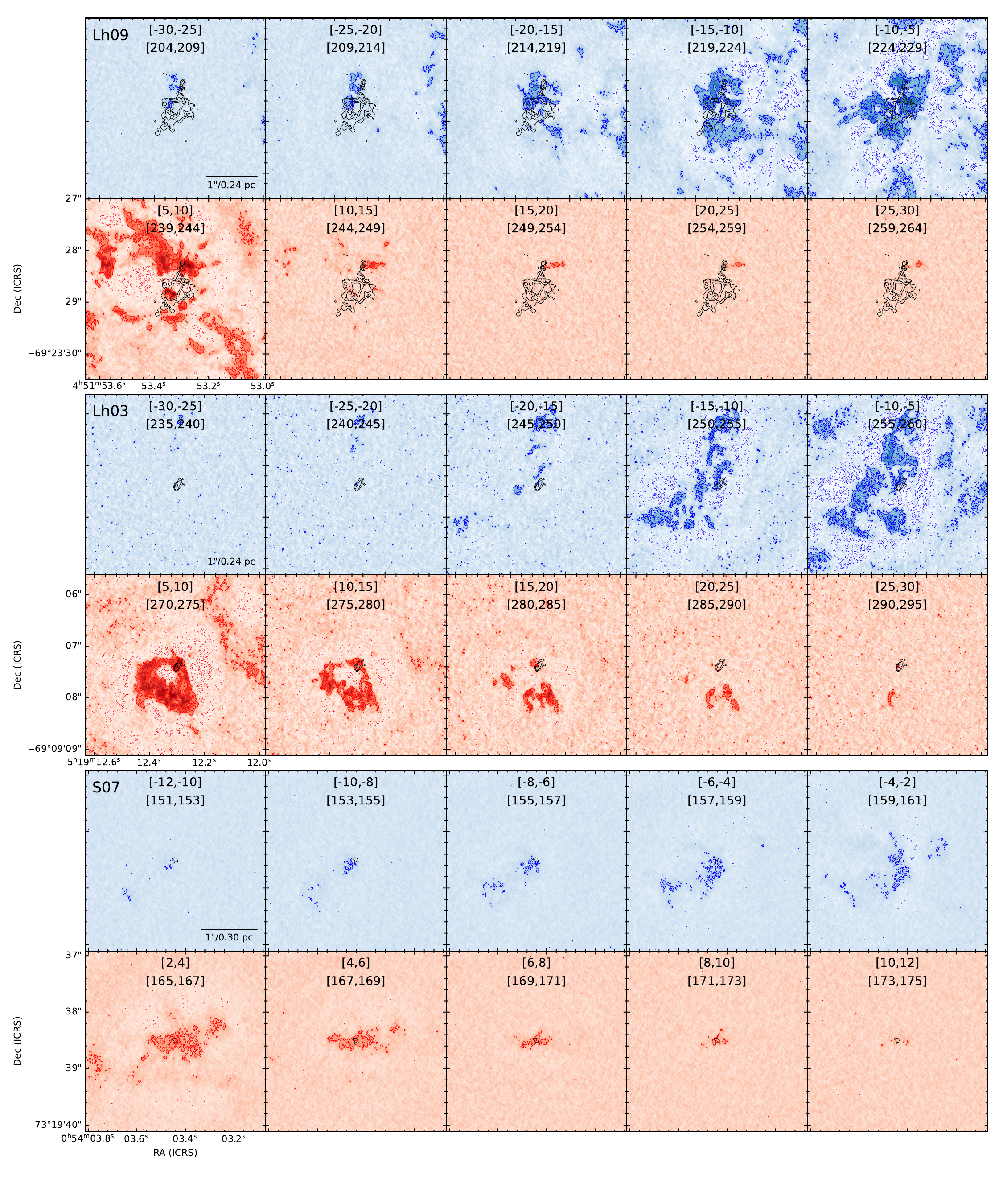}
\caption{CO (3–2) channel maps for Lh09, Lh03, and S07. Each panel is integrated over 5~\kms\ intervals for Lh09 and Lh03, and 2~\kms\ intervals for S07. {The upper velocity interval marked in each panel is relative to the systemic velocity listed in \autoref{table:target_info}, while the lower interval gives the corresponding absolute velocity range.} The CO contours are plotted at (5, 10, 20, 40)$\times\sigma$, where $\sigma = \sqrt{N_{\rm channel}} \times 10^{-3}$\jypbm\kms. The 0.87~mm continuum is overlaid as contours using the same levels as in \autoref{fig:cont}.}
\label{fig:outflow}
\end{figure*} 

\begin{figure*}[ht!]
\centering
\includegraphics[width=1.\textwidth]{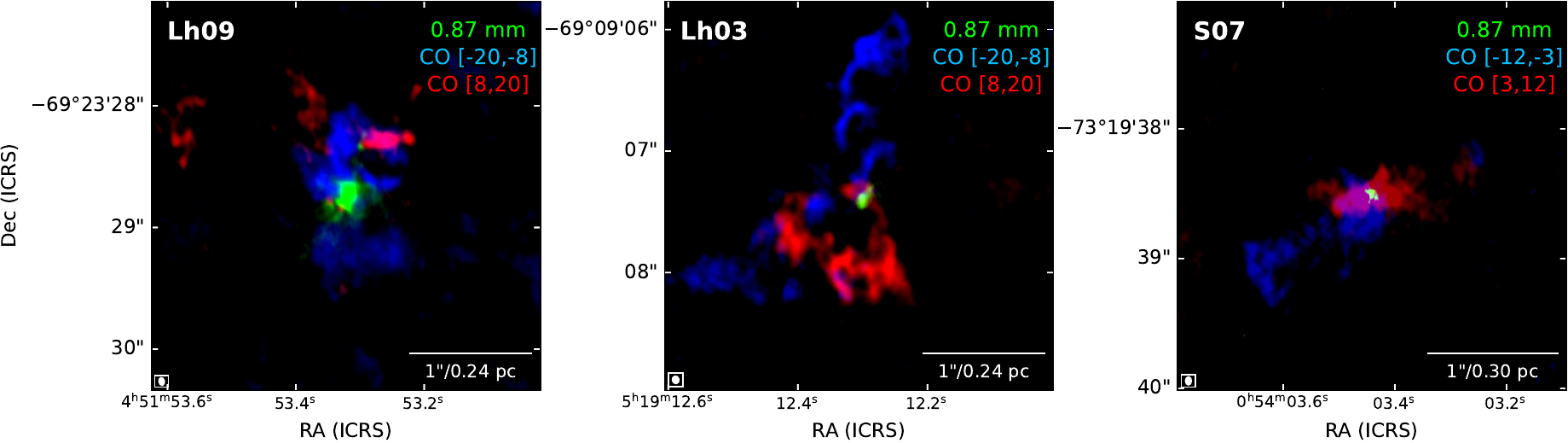}
\caption{Color composite maps of the molecular outflows associated with Lh09, Lh03, and S07. The 0.87~mm continuum emission is shown in green, while CO (3--2) emission integrated over blueshifted and redshifted velocity intervals is shown in blue and red, respectively. The velocity intervals, indicated in each panel, are relative to the systemic velocities listed in \autoref{table:target_info}. The CO data have been spatially smoothed to 1.5 times the native synthesized beam in this plot to improve sensitivity to low surface brightness emission.}
\label{fig:outflow_integrated}
\end{figure*}

\begin{figure*}[ht!]
\centering
\includegraphics[width=1.\textwidth]{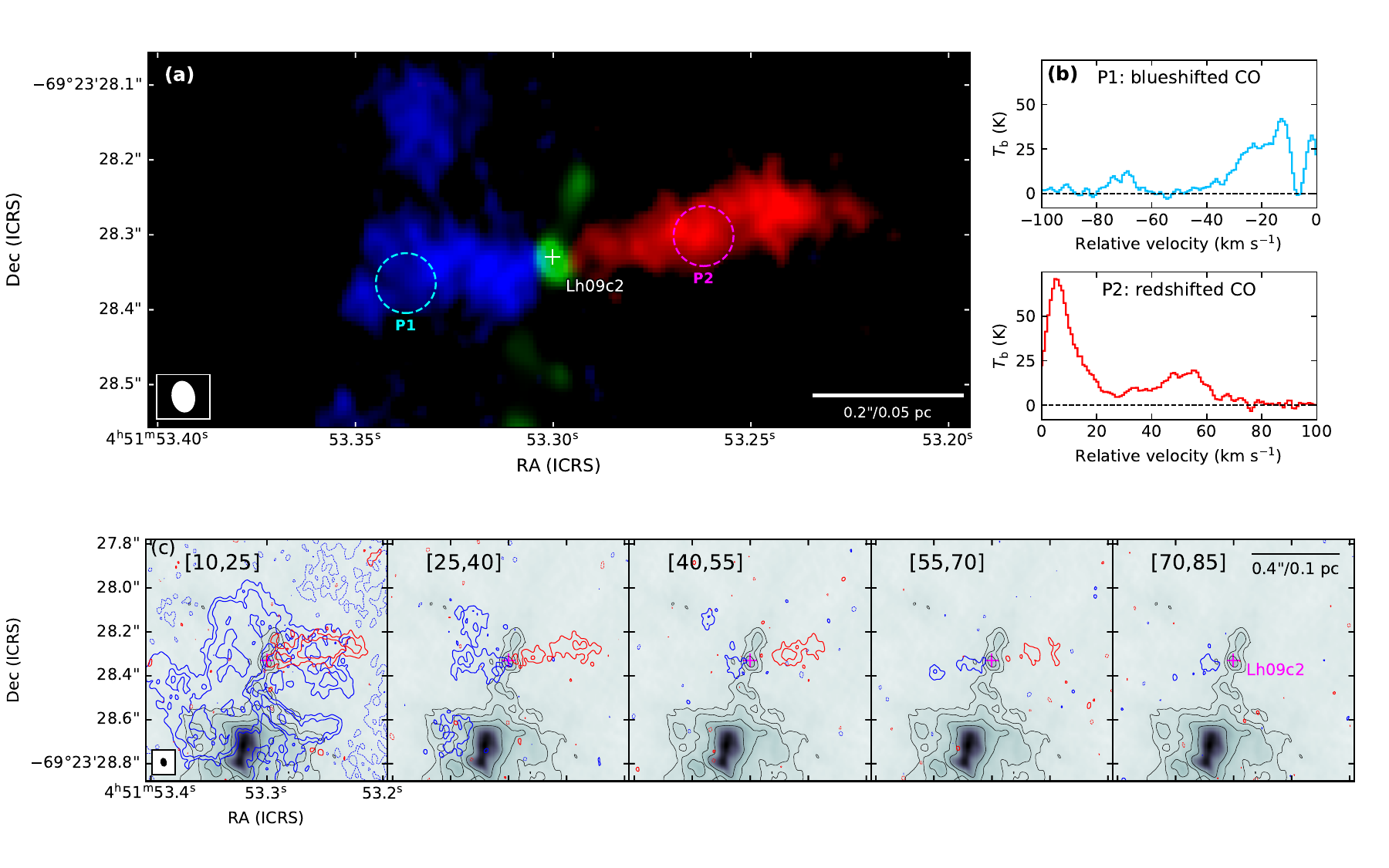}
\caption{A collimated bipolar CO (3--2) outflow associated with Lh09c2. 
{\it (a)} Color composite map of the outflow, with blueshifted and redshifted CO (3--2) emission shown in blue and red, respectively, and the 0.87 mm continuum shown in green. The CO emission is integrated over $-70$ to $-20$ and $+20$ to $+70$ \kms\ relative to the systemic velocity of $V_{\rm sys}=235.1$ \kms, as determined from fitting the \sotwo\ lines. The white cross marks Lh09c2. The cyan and magenta dashed circles mark P1 and P2, respectively, and have radii of $0\farcs04$. 
{\it (b)} Aperture-averaged CO (3--2) spectra toward P1 (upper) and P2 (lower), shown as brightness temperature, $T_{\rm b}$. The spectra are displayed over the blueshifted and redshifted velocity ranges, respectively. The spectra were boxcar-smoothed over three adjacent velocity channels to improve the S/N ratio.
{\it (c)} Channel maps of the bipolar outflow. Each panel shows blue- and redshifted CO (3--2) emission integrated over the indicated velocity interval relative to $V_{\rm sys}$. Blue and red contours show the corresponding blueshifted and redshifted components; solid and dashed contours denote positive and negative emission, respectively. CO contours are plotted at $(4, 8, 16, 32)\times\sigma$, where $\sigma=\sqrt{N_{\rm channel}}\times10^{-3}$ \jypbm\kms. The 0.87 mm continuum is shown in grayscale and black contours at $(5, 10, 20, 40)\times\sigma_{\rm cont}$. }
\label{fig:outflow_Lh09c2}
\end{figure*} 

\begin{figure*}[ht!]
\centering
\includegraphics[width=1.\textwidth]{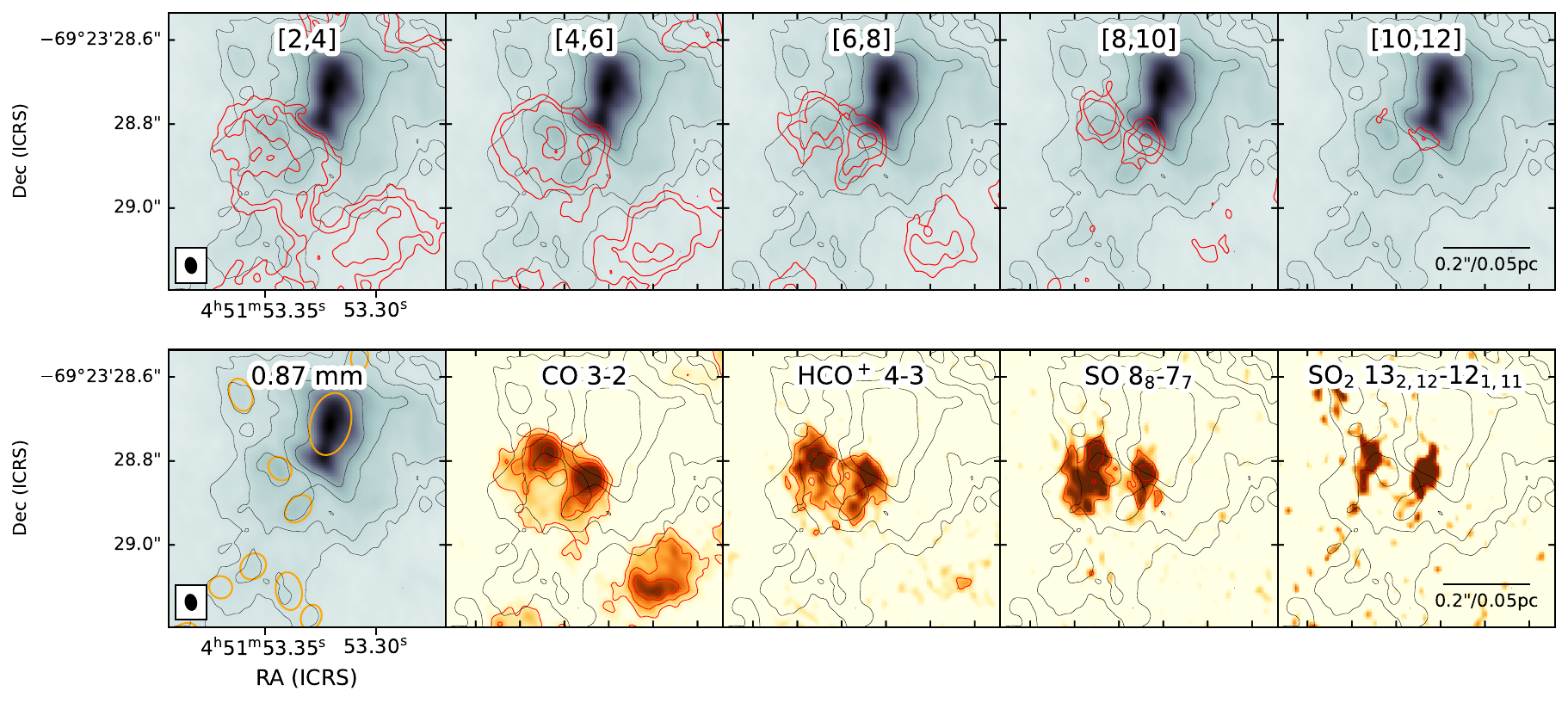}
\caption{A shock-heated outflow to the southeast of Lh09c1. {\it Upper panels:} CO (3–2) integrated intensity maps in 2~\kms\ intervals from 2 to 12~\kms\ (relative to the systemic velocity of 234.1~\kms). CO contours are plotted at (5, 10, 20, 40)$\times\sigma$, where $\sigma = \sqrt{N_{\rm channel}} \times 10^{-3}$\jypbm\kms. The 0.87~mm continuum is overlaid in colorscale and black contours at (5, 10, 20, 40)$\times\sigma_{\rm cont}$. {\it Lower panels:} The leftmost panel shows the 0.87~mm continuum in grayscale and black contours; orange ellipses mark the identified {core candidates}. The remaining panels show integrated intensity maps of CO (3–2), HCO$^+$(4–3), SO~($\rm 9_8-8_7$), \sotwo~($\rm 13_{2,12}-12_{1,11}$), all integrated from 4 to 10~\kms\ relative to the systemic velocity. The line emission is shown in color scale and red contours at (5, 10, 20, 40)$\times\sigma$, where $\sigma = \sqrt{N_{\rm channel}} \times 10^{-3}$\jypbm\kms. The 0.87~mm continuum is overlaid in black contours for reference.}
\label{fig:outflow_knot}
\end{figure*}


{We use the CO(3--2) data to examine the high-velocity gas and protostellar outflows. \autoref{fig:outflow} presents the CO channel maps, while \autoref{fig:outflow_integrated} summarizes the integrated blueshifted and redshifted emission. High velocity outflow emission appears to be present in all three regions. In Lh09 and Lh03, CO emission extends to more than $\sim10$~\kms\ from the systemic velocity and shows spatially coherent structures associated with the central star-forming regions. The outflow activity in Lh09 was also identified previously by \citet{Nayak19} using ALMA CO(2--1) and $^{13}$CO(2--1) observations at a physical resolution of $\sim0.07$~pc. For S07, the line-of-sight velocities are generally lower, with CO emission detected up to approximately $\pm12$~\kms\ from the systemic velocity. A bipolar CO(3--2) outflow was previously identified at $\sim0.1$~pc resolution by \citet{Tokuda22} and \citet{Shimonishi23}, with blueshifted and redshifted wing emission extending over approximately $-16$ to $-5$~\kms\ and $+5$ to $+17$~\kms\ relative to the systemic velocity, respectively. The previously identified bipolar orientation is consistent with that resolved in our higher-resolution observations. Below, we discuss the high velocity CO emission in each target in more detail.}

\subsubsection{Lh09}\label{sec:outflow}

Lh09 exhibits the most complex CO morphology. This is expected, as Lh09 is the most luminous and shows largest number of {compact continuum sources} among the three, indicating the most intense star formation activities. The high velocity gas in the region has been reported in \citet{Nayak19} with CO 2--1 and \thirteenco (2-1) data, but with {relatively} coarse resolution of $\sim$0.07~pc. Although high velocity CO is detected across the 18\arcsec{} FOV, in \autoref{fig:outflow} we focus on the central 4\arcsec{}(corresponding to $\sim$1~pc), where the millimeter continuum is detected, to better associate the CO outflows with their driving sources. At velocities from 5 to 10~\kms{} relative to the systemic velocity of 234.1~\kms, the {CO} morphology is complicated, with substantial emission overlapping and surrounding continuum {sources}. The redshifted lobe is slightly more organized, with some emission extending to the NE and SW with a wide opening angle. Some filamentary or clumpy features are also visible when inspecting the channel maps (see e.g., 246.6~\kms), and some of them are likely outflows associated with individual star forming cores. At higher relative velocities (10--30~\kms), the blueshifted lobe gradually shifts to more spatially concentrated morphology to the north and east of the continuum cores. The redshifed lobe appears to be dominated by a collimated jet-like feature extending horizontally westward from core c2. This structure represents a bipolar outflow, with high-velocity emission beyond the velocity range shown in \autoref{fig:outflow}, as discussed below.

While a full characterization of all outflows in this clustered environment is beyond the scope of this paper, we highlight two particularly interesting prototypical outflows in Lh09.

{\it 1) A collimated bipolar jet from Lh09c2}

In \autoref{fig:outflow_Lh09c2} we show the detection of a well-collimated CO jet originating from Lh09c2. This source is the second brightest {compact continuum source} in Lh09 and exhibits a high gas temperature of $89.2 \pm 5.2$K, as measured from \sotwo\ lines. Although at lower velocities (10–20~\kms\ relative to the systemic velocity) the outflow is confused with other nearby outflows, its bipolar morphology becomes clearly visible at higher velocities ($>$20~\kms), with a blueshifted lobe extending to the east and a redshifted lobe to the west. This outflow shows the highest degree of symmetry and collimation among our sample. It also exhibits exceptionally high velocities, with the blueshifted lobe reaching up to $\sim$80~\kms\ and the redshifted lobe up to $\sim$70~\kms. {The spectra extracted from representative regions in the two lobes (\autoref{fig:outflow_Lh09c2}) show broad outflow wings together with distinct high velocity secondary components.} For the redshifted lobe, which has a projected length of $\sim$0.1pc and a maximum LOS velocity of 70~\kms, we estimate a dynamical timescale of approximately 1400 years. This suggests a very recent ejection event associated with ongoing active star formation in Lh09c2. This is one of the few cases where we can unambiguously identify both the outflow and its driving {source} at $\sim$1500~au scales.

Different from previously reported CO outflows in the Magellanic Clouds \citep{Fukui15,Fukui19,Tokuda19,Tokuda22,Tokuda22b,Nayak19,Shimonishi16,Shimonishi23}, which were generally detected at lower velocities and coarser angular resolutions, the Lh09c2 outflow exhibits a more collimated, jet-like bipolar morphology and reaches higher velocities. These properties resemble the extremely high velocity (EHV) components identified in Galactic protostellar outflow studies, which are characterized by high velocities ($\gtrsim 30~\kms$), jet-like morphologies, and being spectrally detached from the broad outflow wing \citep[e.g.,][]{Bachiller96, Tafalla10, Tychoniec19, Cheng19, Ikeda25}. Such components are often interpreted as tracing material launched from closer to the disk/protostar system, rather than only swept-up ambient gas. {Note that molecular gas directly launched from the protostar--disk system may also contribute at lower velocities \citep[e.g.,][]{Machida13}, whereas the distinct EHV-like component provides a clearer observational signature of such material.} To our knowledge, this is the first spatially resolved molecular CO jet of this kind detected in the Magellanic Clouds, providing strong support for disk-mediated accretion in low-metallicity massive star-forming environments.

{\it 2) Shock excited knots in the outflow}


In \autoref{fig:outflow_knot}, we highlight the redshifted CO feature located immediately to the southeast of the brightest millimeter continuum source, Lh09c1. {At lower velocity offsets ($<$6~\kms), the CO emission shows a slightly conical morphology extending toward the southeast. At higher velocities, it resolves into two compact clumps (each $\sim$0\farcs{1} in size), with line-of-sight velocities up to $\sim$12~\kms. Interestingly, it is also detected in other lines, including HCO$^+$ and the shock tracers SO and SO$_2$. Among our three target regions, SO and SO$_2$ emission typically traces dense gas associated with dust continuum. But in this case, their emission peaks do not coincide with the dust continuum but instead closely match the high velocity CO features, suggesting they may trace shocked material in the outflow. This outflow may be associated with the ejection events from Lh09c1, although contributions from nearby sources cannot be ruled out.}
An LTE analysis of the SO$_2$ lines suggests a high gas temperature of $\sim$85$\pm$5~K. Although we have not detected complex organic molecules in this observation, the outflow gas may have been chemically enriched by shocks, resembling those seen in massive young stellar objects in the Galaxy, such as NGC~7538S \citep{Feng16} and G331.512$-$0.103 \citep{Carlos19}. If confirmed, this source could serve as a valuable laboratory for studying shock chemistry under low-metallicity conditions.

\subsubsection{Lh03 and S07}

As seen in \autoref{fig:outflow}, the most prominent feature for Lh03 is a fan-shaped, wide-angle CO {structure} in the redshifted component, which extends southeastward from the continuum peak, most clearly visible between 5 and 15~\kms\ relative to the systemic velocity. The outflow has an opening angle of approximately 100\arcdeg, with its apex located close to Lh03c3. Portions of this fan-shaped structure persist at higher velocities, up to $\sim$30~\kms. It is unclear whether this high-velocity emission shares the same origin (i.e., driven by the same protostar) as the lower velocity component. On the other hand, the blueshifted lobe is less organized, with relatively diffuse emission extending to the north, east, and south to the continuum sources. Some emission to the north is seen at LOS velocities up to $\sim$30~\kms. At the highest velocities close to 30~\kms, we observe clumpy features roughly symmetrically distributed against the continuum across the red- and blueshifted lobes along a north-south orientation. Overall, the outflows in Lh03 are also likely driven by more than one sources. The complexity of the morphology makes it difficult to isolate individual outflows or confidently identify their driving cores.

S07 exhibits the simplest CO morphology among the sample. It can be described as a bipolar outflow centered on the continuum peak, oriented along the SE–NW direction with a position angle of approximately 120\arcdeg. This outflow was previously reported by \citet{Tokuda22} and \citet{Shimonishi23} using data with $\sim$0.1pc resolution. Our observations, at a resolution of 1800~au, clearly resolve finer structural details. First, in the velocity range of 2 to 6~\kms\ relative to the systemic velocity, a wide-angle component is apparent, forming a conical shell with its vertex approximately at the location of the continuum source. This feature is more clearly seen in the blueshifted lobe. 
Second, although the southeastern lobe is primarily dominated by blueshifted emission, some redshifted emission is also detected at relatively low velocities, particularly in the wide-angle component near the continuum peak. Similarly, redshifted emission is also present in the northwestern lobe. This may be explained if the outflow lies predominantly in the plane of the sky, which would also account for the relatively low LOS velocities in S07 ($\lesssim$12~\kms).

\subsection{Gas Kinematics}\label{sec:kinematics}

\begin{figure*}[ht!]
\centering
\includegraphics[width=1.\textwidth]{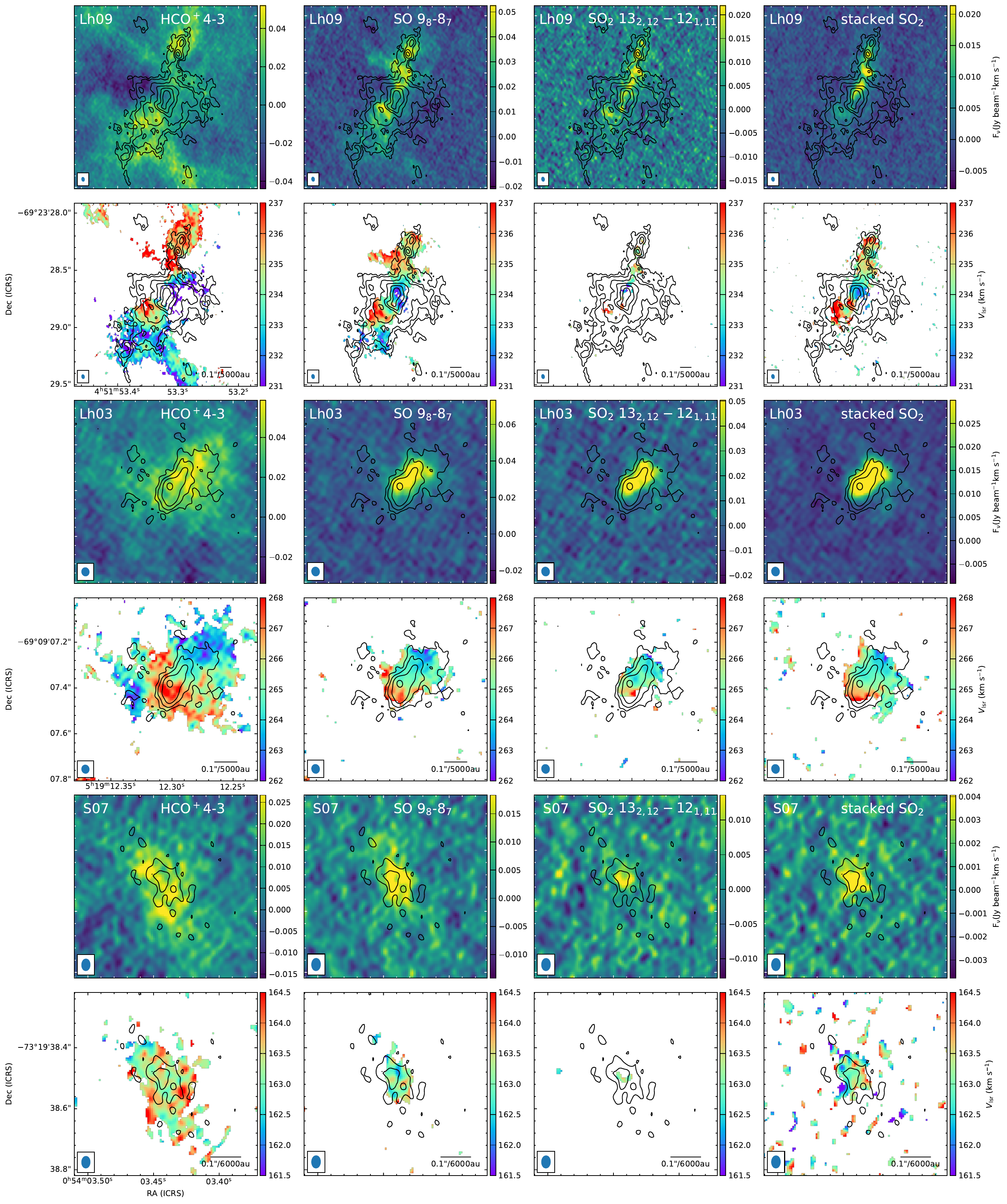}
\caption{Moment 0 and 1 maps for the three targets. From left to right we present the results for \hcop~(4-3), SO~($\rm 9_8-8_7$), \sotwo~($\rm 13_{2,12}-12_{1,11}$), stacked \sotwo, respectively. In all panels the black contours show the 0.87~mm continuum in levels of (3, 6, 12, 25, 50, 100) $\times\sigma_{\rm cont}$ with $\sigma_{\rm cont}$ listed in \autoref{table:target_info}. For moment 1 maps we only show regions with line detection over 5$\sigma$.
}
\label{fig:mom} 
\end{figure*} 

\begin{figure*}[ht!]
\centering
\includegraphics[width=1\textwidth]{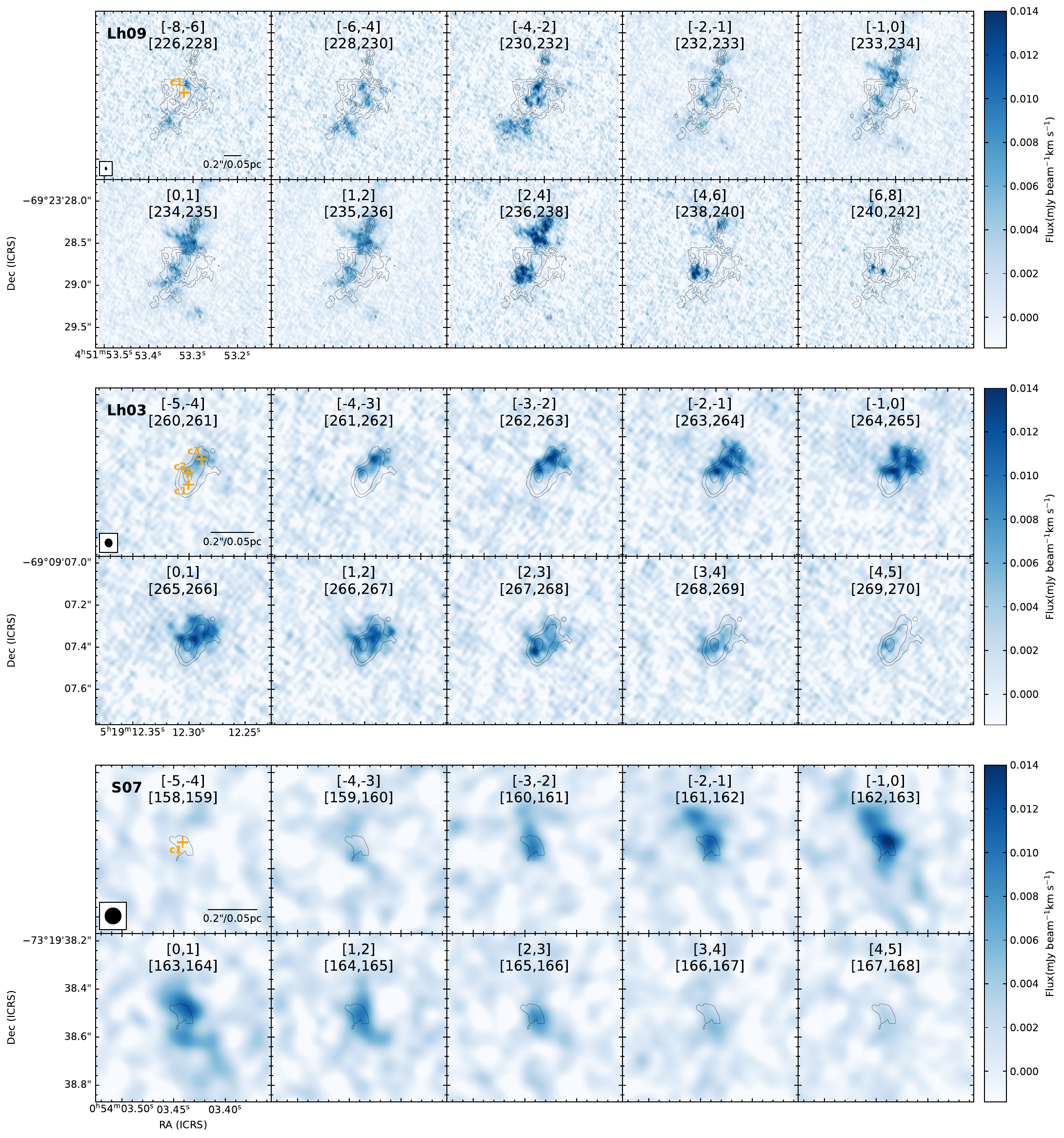}
\caption{Integrated intensity maps of the SO ($9_8$--$8_7$) line for Lh09, Lh03, and S07. The velocity range for each panel is indicated at the top of the panel. Black contours show the 0.87~mm continuum emission at levels of (3, 6, 12, 25, 50, 100)$\times\sigma_{\rm cont}$ for comparison. The orange crosses show the positions of the three core {candidates} in Lh03, Lh09c1 in Lh09 and S07c1 in S07.}
\label{fig:SO_chan} 
\end{figure*} 




\begin{figure*}[ht!]
\centering
\includegraphics[width=1.0\textwidth]{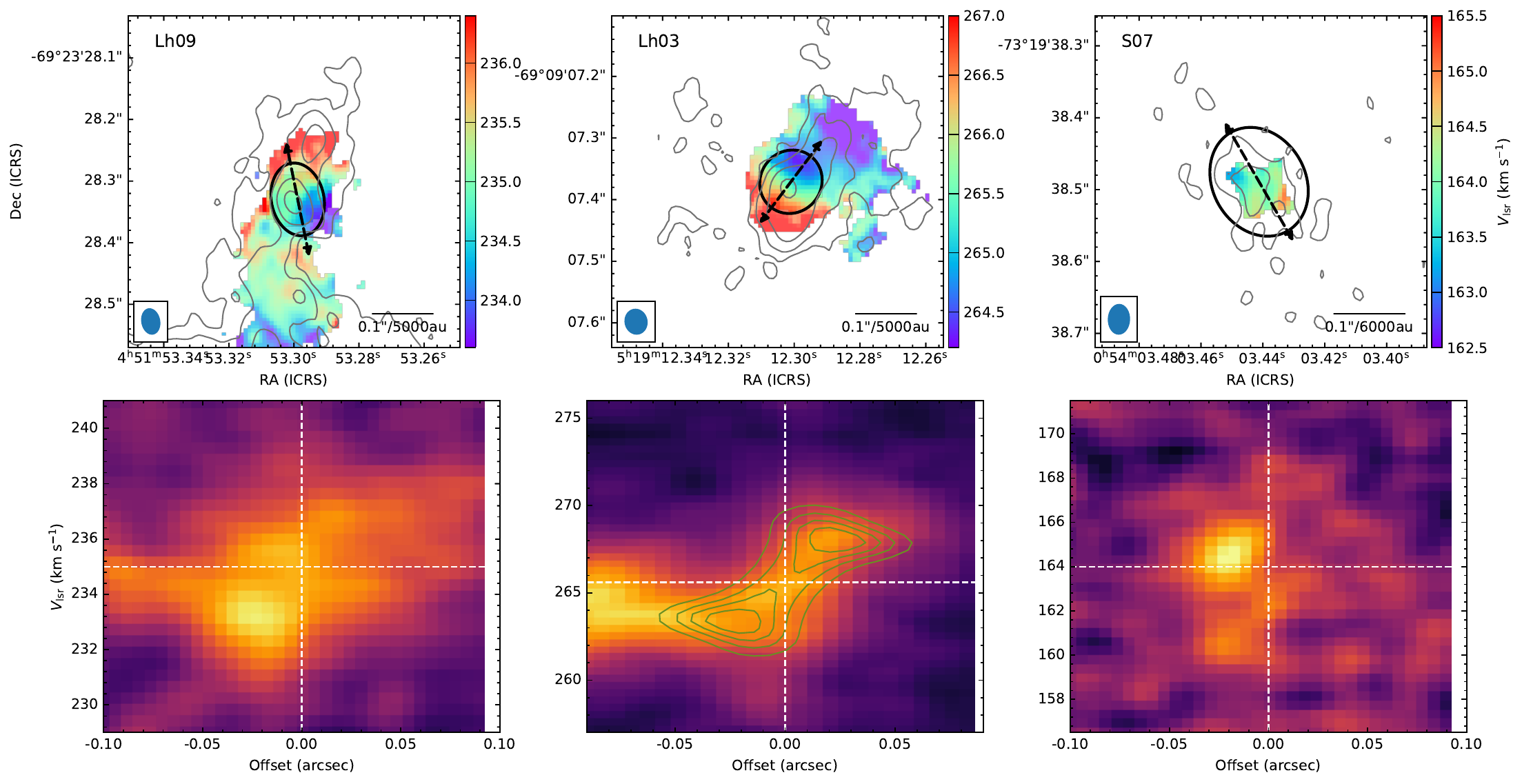}
\caption{{\it (Top)} From left to right, the panels shows moment~1 map of the stacked \sotwo{} line for Lh09c2, Lh03c2, and S07c1.The grey contours show the 0.87~mm continuum emission. The back ellipses mark the corresponding core {candidates} as defined by {\it getsf}, with the major and minor axes set to twice their fitted FWHM values. The dashed black arrow indicate the cut to extract PV diagram shown in bottom panels. {\it (Bottom)} The corresponding position--velocity (PV) diagram extracted from the stacked \sotwo{} cube toward each target. For Lh03c2 The contours show the result of an analytic model for a rotating disk with $m_*$ = 10~\msun, $R_{\rm disk}$ = 3500~au, $R_{\rm in}$ = 1000~au and inclination angle $i$ = 30~\arcdeg.
}
\label{fig:pv_sum} 
\end{figure*} 

In \autoref{fig:mom} we present the moment maps of three targets in \hcop~(4-3), SO~($\rm 9_8-8_7$), \sotwo~($\rm 13_{2,12}-12_{1,11}$). For both Lh09 and Lh03, we detected multiple isolated \sotwo{} lines with comparable strengths. We therefore constructed stacked \sotwo{} line cubes to improve the signal-to-noise ratio for the subsequent kinematic analysis. For this purpose, we selected eight \sotwo{} transitions with upper-state energies in the range $E_{\rm up}=50$--100~K, all of which show similar kinematic structures. For each transition, the data cube was converted to velocity space using the corresponding rest frequency and resampled onto a common velocity grid. The line intensities were then normalized by their integrated line fluxes relative to the reference transition \sotwo{} $13_{2,12}-12_{1,11}$ at 345.3385~GHz, so that all selected transitions were placed on a consistent intensity scale. The normalized cubes were then averaged to produce a stacked \sotwo{} cube. For S07, the detection in each individual \sotwo{} transition is marginal, so we applied the same velocity alignment and stacking procedure, but without normalizing the individual line intensities. The stacked \sotwo{} is also shown \autoref{fig:mom}.

Overall, \hcop~(4-3) traces more extended structures than the 0.87~mm continuum or SO and \sotwo{} lines. The distributions of SO and \sotwo{} are more spatially compact and mostly confined to regions with strong continuum emission. The velocity structures of the three tracers are broadly consistent as seen from the moment~1 maps.

\subsubsection{Lh09}

For Lh09 the line emission is primarily observed along the NW-SE orientated filamentary structure seen in continuum. In \hcop~{} and SO there is also an additional filamentary feature in the NE direction, connecting to this main structure near the position northward of Lh09c1, which is not associated with continuum emission. The \sotwo{} is mainly detected to the north of Lh09c1 along the main filamentary structure (hereafter referred to as the ``north branch''), but some emission to the south is also visible after stacking. Interestingly, for all the lines, there appears to be a deficit of emission at the location of Lh09c1, as well as in the region to the east and west of Lh09c1 despite the presence of extended continuum emission in those areas. This supports the scenario where the material around Lh09c1 has been largely ionized, and the observed 0.87~mm continuum emission (including the E–W extension) is primarily contributed by free-free emission. 

Lh09 displays a complex velocity structure. The north branch is primarily redshifted, with mean velocities around 234–235~\kms, but transitions to $\sim$232~\kms\ near the position of Lh09c1. Southeast of Lh09c1, along the extended continuum emission (hereafter the SE branch), a velocity gradient is observed: the gas is blueshifted ($\sim$231~\kms) near the 5$\sigma$ edge of the continuum emission, becomes redshifted ($\sim$235~\kms) closer to Lh09c1, and then transitions back to blueshifted at the position of Lh09c1. This velocity pattern is most clearly seen in the moment1 map of the SO($9_8$–$8_7$) line.

Since moment 1 maps only reflect the mean velocity distribution, we further illustrate the kinematic complexity in channel maps, with an example shown in \autoref{fig:SO_chan} for the SO~($9_8$–$8_7$) line. The emission spans a broad velocity range from 226 to 242~\kms. It is evident that the velocity gradient seen in \autoref{fig:mom} does not represent a smooth transition. Instead, most regions with detected line emission exhibit a wide velocity span ($>$5~\kms), likely consisting the presence of multiple velocity components arising from different physical processes. One clear example is the contribution from protostellar outflow activity. As discussed in \autoref{sec:outflow}, the redshifted emission associated with the SE branch is most likely tracing a shocked, excited outflow given its close spatial correspondence with high-velocity CO emission, and is reflected in the redshifted components of the HCO$^+$, SO, and \sotwo\ lines. In addition to outflows, other bulk motions within the protocluster, such as gas inflow or rotation, may also contribute to the observed kinematic complexity. A detailed decomposition and analysis of the kinematics in this region will be presented in a future publication.

\subsubsection{Lh03}

As for Lh03, while \hcop{} traces more extended emission, the SO and SO$_2$ lines are primarily detected toward Lh03c2 and Lh03c3, with little to no emission at the position of Lh03c1. This may suggest that Lh03c1 is at a relatively early evolutionary stage and has not yet been sufficiently heated by protostellar radiation. We note that a high temperature of $80.4 \pm 2.9$~K has been reported for Lh03c1 in \autoref{table:core_info}, similar to those of Lh03c2 and Lh03c3. However, there is some overlap in the apertures returned by {\it getsf} between Lh03c1 and Lh03c2, and given that the line emission is clearly centered on Lh03c2, it is most likely that the temperature measurement for Lh03c1 is contaminated by emission from Lh03c2.

A clear velocity gradient is observed, with mean velocities increasing from $\sim$263~\kms\ in the northeast to $\sim$267~\kms\ in the southwest. The channel map of SO~($9_8$–$8_7$) in \autoref{fig:SO_chan} further reveals that the emission can be spatially decomposed into two components associated with Lh03c2 and Lh03c3. The two sources exhibit slightly different mean velocities, and the velocity gradient is more prominent toward Lh03c2, oriented along the NW–SE direction. We further illustrate the velocity gradient in Lh03c2 using the stacked \sotwo{} line in \autoref{fig:pv_sum}, where the velocity range is fine-tuned to highlight the main line emission associated with Lh03c2. A position–velocity (PV) diagram is also shown in \autoref{fig:pv_sum}, extracted along a cut at a position angle of 145\arcdeg. This direction, identified by eye to maximize the observed velocity gradient, is consistent with the elongation of Lh03c2 as returned by {\it getsf}, which has a position angle of 143\arcdeg.

A plausible explanation for the observed velocity gradient in Lh03c2 is a rotational structure associated with the MYSO. This is supported by the fact that the gradient is centered roughly on the continuum peak of Lh03c2. Such rotating structures, sometimes referred to as ``toroids'', can be seen in the velocity fields of high-density tracers in Galactic MYSOs when observed at a few thousand au resolution \citep{Beltran16}. In our case, although the $\sim$1500au resolution is likely insufficient to resolve a genuine Keplerian disk, it is already comparable to the scale of some of the largest massive disks observed in the Galaxy. The estimated mass of Lh03c2 with a temperature of $\sim$80~K, is about 6.9~\msun, which is larger, but comparable to the disk masses found in some MYSO systems \citep{Lu22,McLeod24}. Therefore, the observed velocity gradient may trace a rotating–infalling structure associated with the disk and inner envelope of a single MYSO system.

To further explore this scenario, we apply a simple analytic model described in \citet{Cheng22} to reproduce the observed PV structure, shown as contours in \autoref{fig:pv_sum}. The model assumes an optically thin, uniformly excited rotating disk with a simplified geometric configuration characterized by a vertical height $h(r) = 0.2r$ and a power-law density profile. We tested various parameter combinations, and find that the observed PV diagram can be reasonably reproduced with a model assuming a central stellar mass $m_* = 10$~\msun, an outer radius $R_{\rm disk} = 3500$~au, an inner radius $R_{\rm in} = 1000$~au, and an inclination angle $i = 30\arcdeg$, which corresponds to a nearly edge-on configuration. Given the current coarse spatial and velocity resolution, we refrain from performing a full parameter search or uncertainty analysis, but only present this modeling effort as a validation that a rotation scenario can reasonably explain the observed kinematic signatures.

\subsubsection{S07}

For S07, the line emission is relatively weak.
HCO$^+$(4–3) traces more extended emission and shows stronger intensity toward two blobs located near the northeastern and southwestern tips of the dust continuum. In contrast, SO($9_8$–$8_7$) emission is more compact and closely associated with the continuum. SO$_2$ is marginally detected in the moment~0 map, but it is not sufficiently strong to reveal any kinematic structure. 


We smoothed the spectral data to a resolution of 0\farcs{07} (4200au) to improve the signal-to-noise ratio and present the channel maps of SO~($9_8$–$8_7$) in \autoref{fig:SO_chan}. This plot reveals a blueshifted filamentary emission feature appears northeast of S07c1 from 160 to 164~\kms, connecting to the continuum source S07c1. A similar, though fainter, redshifted feature is seen southwest of S07c1 between 164 and 166~\kms. HCO$^+$ shows a comparable intensity distribution, although with weaker emission near the continuum peak. The origin of these connecting features extending to the northeast and southwest in SO and HCO$^+$ is unclear. We do not find a clear correspondence between these features and the CO outflow shown in \autoref{fig:outflow}. Given that their orientation is roughly aligned with the elongation of the continuum and orthogonal to the CO outflow axis, they may arise from rotation and/or gas inflow potentially feeding the MYSO.

\section{Discussion}\label{sec:discussion}

\subsection{Fragmentation in Lh09}\label{sec:dis_frag}

\begin{figure*}[ht!]
\centering
\includegraphics[width=0.95\textwidth]{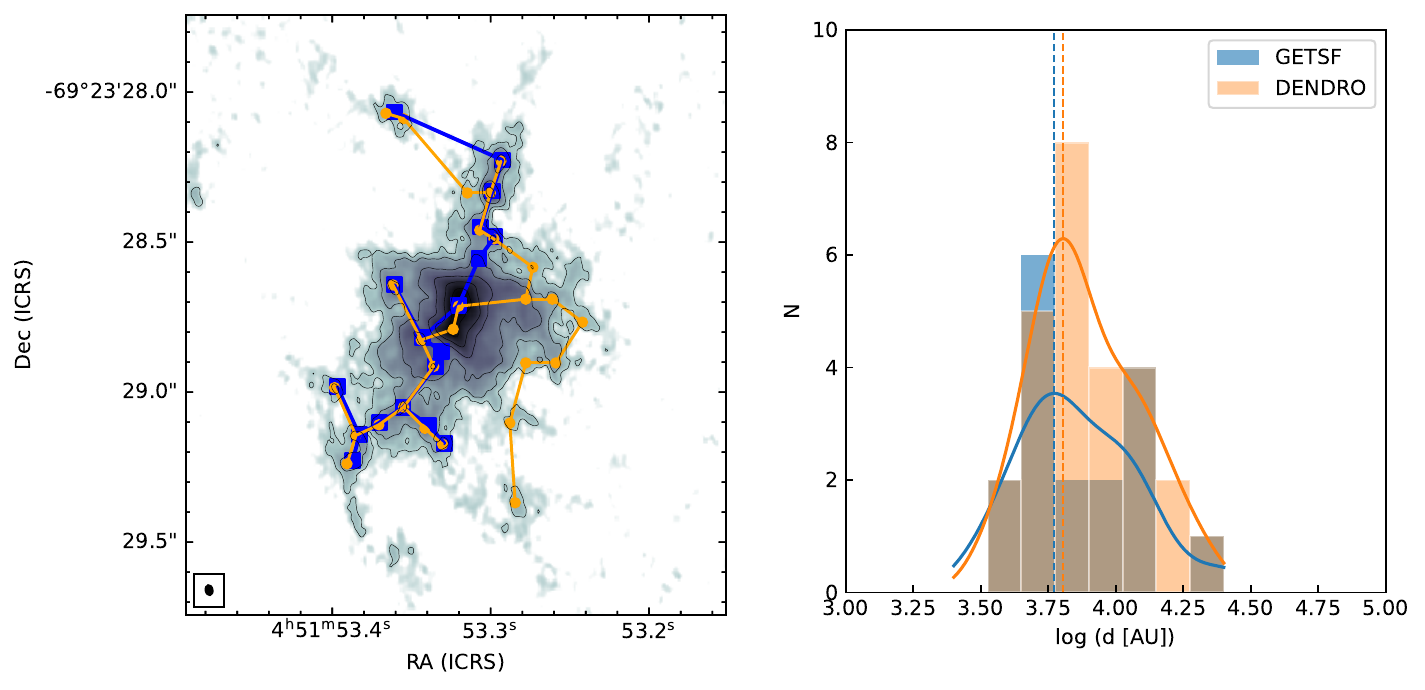}
\caption{ {\it (a)} The colorscale and contours are the same as in \autoref{fig:cont}. The blue segments show the results of the minimum spanning tree, MST, \autoref{sec:dis_frag} using the core identification by {\it getsf}. The orange segments show the results using core identification with {\it astrodendro}. {\it (b)} Distribution of the de-projected separation from the MST analysis for both methods. The curves show the kernel density estimate (KDE) of the distribution, and the dashed vertical lines indicate where the KDE distribution peaks.}
\label{fig:mst} 
\end{figure*} 

Determining the properties of cores, and consequently, the resulting stellar systems—requires an understanding of the dominant mechanisms driving the fragmentation process. In the case of thermal Jeans fragmentation, the characteristic fragmentation scale $\lambda_{J}^{\mathrm{th}}$ is given by $\lambda_{J}^{\mathrm{th}} = c_s (\pi / G \rho)^{1/2}$, where $c_s = (k_B T / \mu m_H)^{1/2}$ is the one-dimensional isothermal sound speed ($k_B$ is the Boltzmann constant, $m_H$ is the mass of a hydrogen atom, and $\mu = 2.37$ is the mean molecular weight per free particle). Here, $G$ is the gravitational constant and $\rho$ is the volume mass density. With ALMA, an increasing number of surveys now achieve spatial resolutions of 1000–2000~au, sufficient to resolve the Jeans length in massive star-forming clumps on scales of a few 0.1~pc with typical density of $10^5--10^{6}$ cm$^{-3}$. These surveys cover a variety of environments (e.g., IRDCs \citep{Sanhueza19, Morii24}, the CMZ \citep{Lu20}, Galactic hot cores \citep{Ishihara24}, etc.), and most of them report a separation scale consistent with thermal Jeans fragmentation, suggesting a potentially universal fragmentation mode. 

Our unprecedented high resolution observations now allow for an examination of the fragmentation properties in the Magellanic Cloud at spatial resolutions comparable to those of Galactic surveys. {Before quantifying the fragmentation, we briefly discuss the nature of the compact continuum sources. For Lh03 and S07, the continuum sources are likely dominated by dust associated with star formation, as supported by their association with dense gas tracers and the CO outflow. The situation in Lh09 is more complex. Lh09c1 is associated with an evolved H\,{\sc ii} region, and its Band~7 continuum emission is dominated by free--free emission (\autoref{sec:free-free}); nearby candidates such as c4 and c10 may likewise be affected by substructure in the ionized emission in addition to dust emission. Lh09c3 is a relatively diffuse continuum feature without a clear molecular line counterpart in the present data. Most of the remaining candidates lie along the north--south filamentary structure traced by HCO$^+$(4--3), and many are also associated with SO or stacked SO$_2$ emission. Among them, Lh09c2 provides particularly clear evidence for an individual star forming source through its well-collimated bipolar EHV-like outflow. Although we cannot confirm that every continuum candidate is a physically distinct structure that is forming, or will form, an individual star, the ensemble nevertheless provides a useful operationally defined population for assessing the fragmentation properties of Lh09.}

To quantify the spatial distribution, we applied the Minimum Spanning Tree (MST) method, which is a graph-based method that connects a set of points with the shortest possible total edge length without forming loops and widely used in studies of spatial clustering \citep[e.g.,][]{Clarke19}. The results are shown in \autoref{fig:mst}a. We also employ another widely used core-identification algorithm, \textit{astrodendro} \citep{Rosolowsky08}, to independently verify the results. We adopt a base flux-density threshold of $4\sigma$, a minimum significance of $1\sigma$, and a minimum area equal to half the synthesized beam area. \textit{astrodendro} appears to provide a more complete identification of local maxima in the continuum map, and we also perform the same MST analysis on the \textit{astrodendro}-identified cores. The distribution of edge lengths is shown in \autoref{fig:mst}b. We multiplied the measured separations by a factor of 4/$\pi$ to account for the projection from 3D to 2D (see Equation 9 in \citealt{Ishihara24}), and the resulting deprojected separation distribution peaks around 6000~au for both methods (5900~au for {\it getsf} and 6300~au for {\it astrodendro}). This separation scale is very similar to the values found in some Galactic surveys \citep[e.g.,][]{Lu20, Xu24, Ishihara24}.

To estimate the thermal Jeans length of the parental clump associated with Lh09, we use the 0.1~pc resolution ALMA Band 7 data from the MAGOS survey \citep{Shimonishi26}, and estimate the mass of the parental clump within a radius of $\sim$0.2~pc, which encompasses the identified core {candidates} studied here. After accounting for the contribution of free-free emission (\autoref{sec:free-free}), and assuming optically thin thermal dust emission at a temperature of 60~K (based on the spectral energy distribution (SED) analysis in \citealt{Ochsendorf17}), we derive a clump mass of $\sim$1300\msun{} using the dust opacity assumptions described in \autoref{sec:cont}. This yields a thermal Jeans fragmentation scale of $\sim$13000~au assuming a same gas temperature of $\sim$60~K. 

The observed core separations in Lh09 are thus comparable but smaller than the thermal Jeans scale estimated from the present clump properties. {Given our observations are less mass sensitive than Galactic surveys, deeper observations would likely reveal additional low-mass cores, increasing the core number density and further reducing the measured mean separation. On the other hand, our observations were obtained with a single long-baseline configuration, and spatial filtering and incomplete $uv$ sampling may introduce spurious substructures in intrinsically smooth emission, particularly for low-contrast features \citep[e.g.,][]{Caselli19,Tokuda20}. To examine this, we repeat the MST analysis using the 12 {\it getsf} sources with local peak intensities greater than 6 times the global image rms. This conservative selection removes several relatively diffuse or low-significance candidates such as c3, c16, and c17. The resulting de-projected separation distribution peaks at $\sim8000$~au, somewhat larger than the fiducial value of $\sim6000$~au but still smaller than the estimated thermal Jeans length. Thus, the exact peak separation is somewhat sensitive to source selection, but the qualitative comparison with the clump-scale thermal Jeans length remains unchanged.} Such a small $\lambda_{\rm sep}/\lambda_{\rm J}$ ratio is also seen in some relatively evolved Galactic clumps \citep{Yang25}. This may reflect the evolved nature of Lh09, as also indicated by its strong ionizing emission (\autoref{sec:free-free}). As a result, its present-day core separations may have been reduced by global contraction and/or continued hierarchical fragmentation after the initial fragmentation stage \citep{Xu24}. In this case, the current clump averaged temperature and density may no longer directly trace the conditions under which the cores first formed. Alternatively, the relevant fragmentation scale may be set by denser local substructures rather than by the clump averaged density, which can also yield smaller core separations than the Jeans length inferred from the larger scale clump properties.


\subsection{Disk rotation for MYSO in the Magellanic Clouds?}

Channeling material through a circumstellar accretion disk is known to be an effective mechanism for overcoming powerful radiative feedback during the formation of massive stars \citep{Krumholz09, Rosen16}. Observationally, it is still not fully established whether circumstellar Keplerian disks form around stars of all masses. In recent years, ALMA observations have gradually built up evidence for this paradigm, and Keplerian disks associated with proto-B and proto-O stars have been reported \citep[e.g.,][]{Johnston15, Ilee16, Sanna19, Lu22,Olguin26,Yang26}. What is even less explored is whether this mechanism is applicable across different cosmic environments. The Magellanic Clouds are among the few, if not the only, extragalactic targets where detailed observations of individual MYSOs can be conducted to test this. In particular, their low metallicity provides a local analog to the conditions in the early universe or high-redshift galaxies, offering useful insight into massive star formation in a broader cosmological context.

The best-characterized MYSO to date in the Magellanic Clouds is HH~1177, which has been identified as a massive protostar system with a collimated jet and a rotating toroid \citep{McLeod18, McLeod24}, consistent with the canonical picture of disk-mediated star formation. Unlike most Galactic counterparts, HH1177 is optically revealed, likely having cleared out much of its natal material. While this facilitates characterization of its intermediate circumstellar dynamics, it also makes HH~1177 less representative of typical MYSOs, which is expected to evolve within deeply embedded protocluster environments during their main accretion phase.

In this study, the outflow and kinematic signatures in the targeted regions offer new insight for the MYSO in the Magellanic Clouds. Lh09c2 exhibits a well-collimated, symmetric bipolar jet reaching terminal velocities of $\sim$80 km/s, which is a clear signpost of ongoing disk-driven accretion. In contrast, Lh03c2 shows no prominent outflow but presents a clear velocity gradient across Lh03c2 that can be described by a rotating toroidal structure feeding a massive protostar. S07c1 displays a symmetric wide-angle bipolar outflow but lacks a clear velocity gradient on $\sim$2000~au scale, though tentative gradient may trace inflowing material. To search for potential disk rotation, we further present the moment~1 maps of the stacked \sotwo{} emission for Lh09c2 and S07c1 in \autoref{fig:pv_sum}. For Lh09c2, we extracted the PV diagram along a position angle of 11\arcdeg, corresponding to the major axis returned by \textit{getsf}; this direction is also approximately perpendicular to the outflow axis. For S07c1, the continuum elongation is less well constrained because of the weak detection, and we therefore adopted a position angle of 30\arcdeg, roughly following the continuum morphology and approximately perpendicular to the outflow direction. However, we do not find a clear velocity gradient consistent with rotation in either source. For Lh09c2, a weak velocity gradient may be present, but it is likely mixed with, or confused by, the kinematics of the larger north--south filamentary structure. In the case of S07c1, the high-excitation SO$_2$ lines are only marginally detected as point sources; this may imply that the hot disk presence is relatively compact and has limited influence on the kinematics at 2000~au scales. 

Overall, the presence of collimated bipolar jets and ordered rotation in these sources supports the notion that in such low-metallicity conditions massive stars are built via disk accretion as in the Milky Way. Nevertheless, we have not yet identified an unambiguous case that simultaneously exhibits both rotational motion and a collimated jet. This is not surprising given the complexity inherent in protocluster environments, which can complicate the detection of outflows and circumstellar kinematics, as may be the case for Lh09c2 and Lh03c2. Furthermore, Galactic MYSOs themselves exhibit a diversity of structures and kinematic behaviors on sub-1000~au scales, including clear envelope--disk transitions \citep[e.g.,][]{Zhang19,Yang26}, complex streamer-like inflows \citep[e.g.,][]{Goddi20}, and hierarchical fragmentation without prominent disk structures \citep[e.g.,][]{Beuther19}. This complexity is also seen in the recent DIHCA survey, where many Galactic high-mass disk candidates show asymmetric PV structures rather than simple Keplerian patterns \citep{Olguin26}. The detection of large Keplerian disks is therefore not always expected and may depend on individual source properties such as the evolutionary stage \citep{Cesaroni17}. Future observations toward a larger sample, ideally with even higher sensitivity and angular resolution, will be essential for further exploring the formation of massive stars in the Magellanic Clouds.

\section{Conclusion}\label{sec:conclusion}

We present the highest resolution ($\sim$1500~au) ALMA Band~7 observations to date of three massive star formation sites in the Magellanic Clouds, Lh03 and Lh09 in the LMC, and S07 in the SMC. The data reveal a diverse range of fragmentation, kinematic, and outflow properties, offering new insights into how massive stars form in such low metallicity environments. Our main conclusions are as follows:

\begin{itemize}

\item {The three systems exhibit different continuum morphologies and fragmentation properties. Lh09 shows extensive fragmentation, with 17 compact continuum sources distributed within the central $\sim0.4$~pc region, whereas Lh03 and S07 show more limited fragmentation. The characteristic de-projected separations in Lh09 peak around $\sim6000$~au, smaller than the thermal Jeans length estimated from the parental clump properties. }

\item  {All targets exhibit molecular outflows traced by high velocity CO 3-2 emission. We identify a well-collimated, symmetric bipolar jet with terminal LOS velocities up to 80~\kms{} driven by Lh09c2, as well as compact molecular knots seen in SO and \sotwo{} that may trace shocked gas associated with outflow activity near Lh09c1.}

\item  {Spectral lines from dense gas tracers such as HCO$^+$, SO, and SO$_2$ are detected in all three targets. The gas kinematics in Lh09 are complex, with multiple overlapping components revealed in channel and moment maps. Velocity gradients and broad line widths are observed across the filamentary structures, likely tracing a mixture of inflow, outflow, and feedback from forming massive stars. In Lh03c2, we detect a clear velocity gradient consistent with a rotating toroid. In contrast, no clear rotation like velocity gradients are detected toward Lh09c2 or S07c1, despite their association with prominent bipolar outflows.}
    
\end{itemize}

Taken together, these findings, including the collimated jets and rotating toroids, are qualitatively similar to what is seen in Galactic MYSOs and we do not find evidence for significant different processes in the formation of massive stars in the low metallicity Magellanic Clouds. Meanwhile, the source-to-source diversity observed across this small sample highlights the need for larger surveys to robustly assess the protostellar evolution of MYSOs, and to uncover any subtle metallicity-dependent trends.

\acknowledgments
This paper makes use of the following ALMA data: ADS/JAO.ALMA\#2023.1.01629.S. ALMA is a partnership of ESO (representing its member states), NSF (USA) and NINS (Japan), together with NRC (Canada), MOST and ASIAA (Taiwan), and KASI (Republic of Korea), in cooperation with the Republic of Chile. The Joint ALMA Observatory is operated by ESO, AUI/NRAO and NAOJ. The National Radio Astronomy Observatory is a facility of the National Science Foundation operated under cooperative agreement by Associated Universities, Inc. Y.C. was partially supported by Grant-in-Aid for Scientific Research (KAKENHI  number JP24K17103 and 26K00748) of the JSPS. K.T. and S.Z. acknowledge support from the NAOJ ALMA Scientific Research Grant Code 2025-29B. Data analysis was in part carried out on the Multi-wavelength Data Analysis System operated by the Astronomy Data Center (ADC), National Astronomical Observatory of Japan.

\vspace{10mm}
\facilities{Atacama Large Millimiter/submillimeter Array (ALMA), Hubble Space Telescope (HST)}
\software{CASA \citep{McMullin07}, APLpy \citep{Robitaille12}, Astropy \citep{Astro13}}

\clearpage

\appendix
\counterwithin{figure}{section}
\counterwithin{table}{section}

\section{Properties of {compact continuum sources}}\label{sec:core_prop}

In \autoref{table:core_info} we list the derived properties for the core candidates in the three targets.
\startlongtable
\begin{deluxetable}{cccccccccc}
\tabletypesize{\scriptsize}
\renewcommand{\arraystretch}{1.0}
\tablecaption{Properties of core candidates in three targets in Magellanic clouds \label{table:core_info}}
\tablehead{
\colhead{Region} & \colhead{Index} & \colhead{$\alpha$(J2000)} & \colhead{$\delta$(J2000)} & \colhead{Flux} & \colhead{Peak Flux Density} & \colhead{$\rm FWHM$} & \colhead{PA} & \colhead{$T_{\rm rot}$} & \colhead{$\rm Mass$} \\
\colhead{} & \colhead{} & \colhead{hh:mm:ss} & \colhead{dd:mm:ss} & \colhead{mJy} & \colhead{mJy beam$^{-1}$} & \colhead{mas$\times$mas} & \colhead{$^{\circ}$} & \colhead{K} & \colhead{$M_{\odot}$}
}
\startdata
S07 & 1 & 00:54:03.441 & $-$73:19:38.49 & 0.58 & 0.17 & 72$\times$59 & 65 & - & 17.6 \\
Lh03 & 1 & 05:19:12.300 & $-$69:09:07.43 & 1.49 & 0.51 & 71$\times$66 & 163 & 80.4(2.9) & 9.2 \\
Lh03 & 2 & 05:19:12.301 & $-$69:09:07.37 & 1.12 & 0.77 & 48$\times$45 & 143 & 80.4(2.8) & 6.9 \\
Lh03 & 3 & 05:19:12.289 & $-$69:09:07.31 & 0.97 & 0.31 & 98$\times$60 & 158 & 83.5(2.6) & 5.7 \\
Lh09 & 1 & 04:51:53.320 & $-$69:23:28.71 & 48.05 & 5.56 & 151$\times$96 & 166 & 96.1(4.0) & -$^{c}$ \\
Lh09 & 2 & 04:51:53.299 & $-$69:23:28.33 & 2.23 & 1.29 & 54$\times$39 & 11 & 89.2(5.2) & 12.2 \\
Lh09 & 3 & 04:51:53.361 & $-$69:23:28.07 & 1.76 & 0.11 & 175$\times$148 & 31 & - & 18.6 \\
Lh09 & 4 & 04:51:53.343 & $-$69:23:28.82 & 1.33 & 0.63 & 62$\times$50 & 42 & - & 14.0 \\
Lh09 & 5 & 04:51:53.335 & $-$69:23:28.91 & 1.27 & 0.55 & 74$\times$53 & 137 & - & 13.4 \\
Lh09 & 6 & 04:51:53.293 & $-$69:23:28.23 & 1.23 & 0.61 & 67$\times$44 & 168 & - & 13.0 \\
Lh09 & 7 & 04:51:53.307 & $-$69:23:28.45 & 0.99 & 0.35 & 75$\times$55 & 4 & 68.5(2.9) & 7.3 \\
Lh09 & 8 & 04:51:53.387 & $-$69:23:29.23 & 0.97 & 0.28 & 92$\times$68 & 140 & - & 10.2 \\
Lh09 & 9 & 04:51:53.339 & $-$69:23:29.11 & 0.94 & 0.28 & 91$\times$62 & 10 & - & 9.9 \\
Lh09 & 10 & 04:51:53.361 & $-$69:23:28.64 & 0.74 & 0.32 & 80$\times$53 & 20 & - & 7.8 \\
Lh09 & 11 & 04:51:53.355 & $-$69:23:29.05 & 0.68 & 0.29 & 66$\times$57 & 138 & - & 7.2 \\
Lh09 & 12 & 04:51:53.307 & $-$69:23:28.55 & 0.47 & 0.32 & 63$\times$41 & 175 & 88.6(5.4) & 2.6 \\
Lh09 & 13 & 04:51:53.297 & $-$69:23:28.48 & 0.43 & 0.41 & 45$\times$33 & 25 & 75.0(3.7) & 2.9 \\
Lh09 & 14 & 04:51:53.370 & $-$69:23:29.10 & 0.35 & 0.20 & 56$\times$53 & 68 & - & 3.7 \\
Lh09 & 15 & 04:51:53.329 & $-$69:23:29.17 & 0.30 & 0.17 & 61$\times$48 & 158 & - & 3.2 \\
Lh09 & 16 & 04:51:53.383 & $-$69:23:29.14 & 0.28 & 0.19 & 66$\times$37 & 47 & - & 3.0 \\
Lh09 & 17 & 04:51:53.397 & $-$69:23:28.98 & 0.25 & 0.22 & 44$\times$34 & 33 & - & 2.6 \\
\enddata
\tablenotetext{a}{Excitation temperatures (and associated uncertainties in brackets) derived from fitting multiple \sotwo{} lines (see \autoref{sec:app_temperature}).}
\tablenotetext{b}{Mass derived from the 0.87~mm continuum emission assuming optically thin thermal dust emission. For sources without temperature measurements we assume 50~K.}
\tablenotetext{c}{The Band 7 continuum of Lh09c1 is dominated by free-free emission (see \autoref{sec:free-free}).}
\end{deluxetable}

\onecolumngrid
\section{Temperature estimation with \sotwo{} lines}\label{sec:app_temperature}

\begin{figure}[ht!]
\centering
\includegraphics[width=1.0\textwidth]{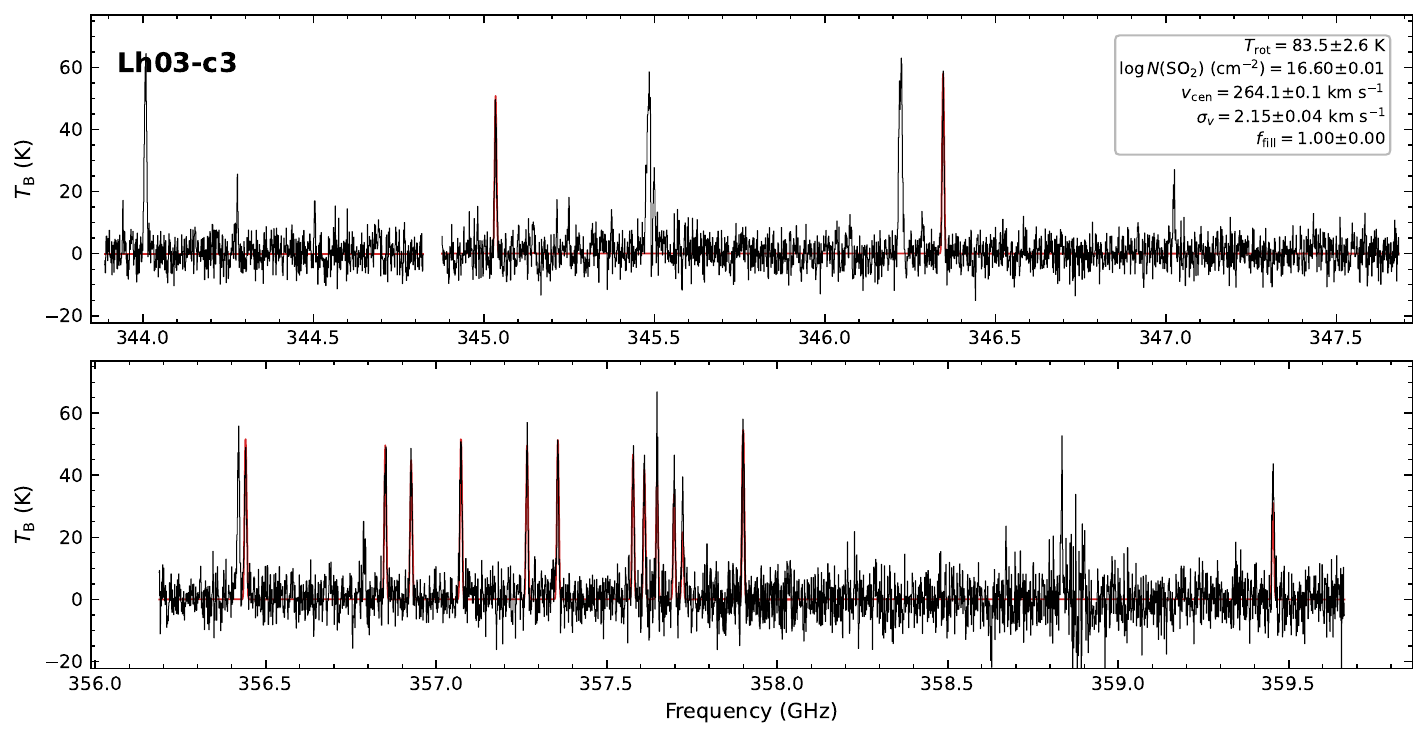}
\caption{ALMA spectra toward Lh03c3 used for the SO$_2$ excitation analysis.
The black step histograms show the extracted brightness-temperature spectra, and the red curves show the best-fit LTE SO$_2$ model with the beam filling factor fixed to unity.
The upper and lower panels cover the observed SO$_2$ transitions near 344--348 GHz and 356--360 GHz, respectively.
}
\label{fig:spec_example}
\end{figure}

We estimate the excitation temperatures of the dense core {candidates}
using multiple SO$_2$ transitions covered by the Band~7 spectral
windows. For each continuum {source}, we extracted the mean spectrum within the
elliptical FWHM footprint returned by {\it getsf}.  The extraction was
performed on the primary-beam-corrected spectral cubes. 
Spectra from all spectral windows containing SO$_2$
transitions (except for \sotwo~$16_{4,12}-16_{3,13}$ around 346.52~GHz that is blended with SO~$9_8-8_7$) were then fitted simultaneously.

We assume local thermodynamic equilibrium (LTE), so that the level
populations are described by a single excitation temperature
$T_{\rm ex}$.  For a transition $u\rightarrow l$, the upper-level column
density is
\begin{equation}
N_u =
N_{\rm SO_2}
\frac{g_u \exp(-E_u/kT_{\rm ex})}
{Q(T_{\rm ex})},
\end{equation}
where $N_{\rm SO_2}$ is the total SO$_2$ column density, $g_u$ is the
upper-level degeneracy, $E_u$ is the upper level energy, and
$Q(T_{\rm ex})$ is the partition function. The spectroscopic
parameters, including rest frequency, Einstein $A$ coefficient,
upper-state energy, degeneracy, and partition function, were taken from
the CDMS/JPL line catalogs through Splatalogue
\citep{Pickett98,Muller01,Muller05}.

For each transition, the optical-depth profile is modeled as
\begin{equation}
\tau_i(v) =
\frac{c^3}{8\pi\nu_i^3}
A_{ul,i} N_{u,i}
\left[\exp\left(\frac{h\nu_i}{kT_{\rm ex}}\right)-1\right]
\phi_i(v),
\end{equation}
where $\nu_i$ is the rest frequency, $A_{ul,i}$ is the Einstein
coefficient, and $\phi_i(v)$ is a normalized Gaussian line profile,
\begin{equation}
\phi_i(v) =
\frac{1}{\sqrt{2\pi}\sigma_v}
\exp\left[-\frac{(v-v_{\rm LSR})^2}{2\sigma_v^2}\right],
\end{equation}
with $\int \phi_i(v)\,dv=1$.  The model brightness temperature is then
\begin{equation}
T_{{\rm b},i}(v) =
\eta_{\rm bf}
\left[J_{\nu_i}(T_{\rm ex}) - J_{\nu_i}(T_{\rm bg})\right]
\left[1-\exp[-\tau_i(v)]\right],
\end{equation}
where
\begin{equation}
J_\nu(T) =
\frac{h\nu/k}{\exp(h\nu/kT)-1},
\end{equation}
$T_{\rm bg}=2.73$ K, and $\eta_{\rm bf}$ is the beam filling factor.
In the fiducial fits we adopt a uniform filling factor of 1 for all SO$_2$ transitions of a given core.
The free parameters in the LTE fit are therefore the systemic velocity
$v_{\rm LSR}$, excitation temperature $T_{\rm ex}$, SO$_2$ column density
$N_{\rm SO_2}$, and the velocity dispersion $\sigma_v$.  The best-fit parameters were found
by minimizing
\begin{equation}
\chi^2 =
\sum_j
\left[
\frac{T_{{\rm obs},j}-T_{{\rm model},j}}
{\sigma_j}
\right]^2,
\end{equation}
where $\sigma_j$ is the rms noise measured from line-free channels of the mean spectra. \autoref{fig:spec_example} shows an example for the fit. The quoted uncertainties correspond to the formal $1\sigma$ errors from the covariance matrix of the nonlinear least-squares fit.

\section{Free-free emission in Lh09}\label{sec:free-free}

\begin{figure}[ht!]
\centering
\includegraphics[width=1.0\textwidth]{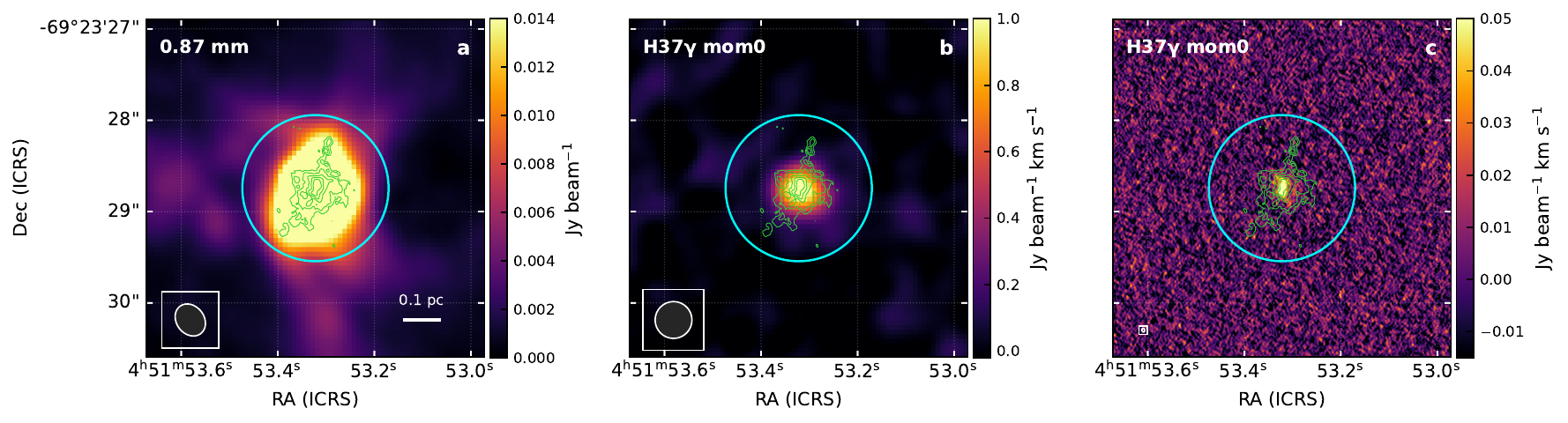}
\caption{
ALMA 0.87 mm continuum and H37$\gamma$ emission toward Lh09.
{\it (a)} 0.1~pc resolution 0.87~mm continuum image.
{\it (b)} 0.1~pc resolution H37$\gamma$ moment 0 map integrated over 210--270 km s$^{-1}$.
{\it (c)} High-resolution (1500~au) H37$\gamma$ moment 0 map over the same velocity range.
In all panels green contours show the high resolution 0.87 mm continuum at 5, 10, 20, 40, and 80$\sigma$.
The cyan circle marks a radius of 0.8$^{\prime\prime}$ centered on Lh09c1.
}
\label{fig:h37g}
\end{figure} 

We estimate the free-free contribution to the 0.87~mm continuum based on the
H37$\gamma$ hydrogen recombination line included in the spectral setup. For optically thin ionized gas,
the ratio between the velocity integrated
recombination line flux density and the free-free continuum flux density can be described as \citep{Bendo15},
\begin{equation}
\frac{\int S_{\nu}({\rm line})\,dv}{S_{\nu}^{\rm ff}}
=
5.06\times10^{32}
\left(\frac{\epsilon_{\nu}}{\rm erg~s^{-1}~cm^{3}}\right)
\left(\frac{\nu}{\rm GHz}\right)^{-0.83}
\left(\frac{T_{\rm e}}{\rm K}\right)^{1/2}
~{\rm km~s^{-1}},
\label{eq:rrl_ff_bendo}
\end{equation}
where $\epsilon_{\nu}$ is the recombination line emissivity per unit
$n_{\rm e}n_{\rm p}$, $\nu$ is the line rest frequency, and $T_{\rm e}$ is
the electron temperature. The weak frequency dependence of the free-free
Gaunt factor is included in the $\nu^{-0.83}$ term. For H37$\gamma$, the transition is $n_{\rm u}=40\rightarrow n_{\rm l}=37$, with $\nu_0=346.76$ GHz. We adopt the Case-B hydrogen emissivity from the recombination-line calculations of \citet{Storey95}.  For $T_{\rm e}=10^4$ K and $n_{\rm e}=10^4~{\rm cm^{-3}}$, the Storey \& Hummer
table gives $\epsilon_{\nu}({\rm H37}\gamma)=3.055\times10^{-32}~{\rm erg~s^{-1}~cm^{3}}$. Substituting this emissivity into \autoref{eq:rrl_ff_bendo} gives
\begin{equation}
S_{\nu}^{\rm ff}(0.87{\rm~mm})
\simeq
0.083
\left[
\frac{\int S_{\nu}({\rm H37}\gamma)\,dv}
{\rm Jy~km~s^{-1}}
\right]
{\rm Jy}.
\end{equation}
If singly ionized helium contributes to the free-free continuum while the
line traces only hydrogen recombination, the continuum estimate should be
multiplied by $(1+y^+)$, where
$y^+=n({\rm He^+})/n({\rm H^+})$. Adopting $y^+=0.08$ gives
\begin{equation}
S_{\nu}^{\rm ff}(0.87{\rm~mm})
\simeq
0.090
\left[
\frac{\int S_{\nu}({\rm H37}\gamma)\,dv}
{\rm Jy~km~s^{-1}}
\right]
{\rm Jy}\label{eq:h37g_ff}
\end{equation}

In \autoref{fig:h37g}, we show the H37$\gamma$ moment map integrated over 210--270~\kms. In the 1500~au resolution data, the H37$\gamma$ emission is mainly concentrated within the central 0\farcs{25}, which is most likely associated with Lh09c1 and its immediate surroundings. The integrated line flux is 1.65~Jy~\kms, corresponding to a free-free continuum flux density of 0.148~Jy following \autoref{eq:h37g_ff}. This is very similar to the measured 0.87~mm flux density within this aperture, 0.144~Jy, and much higher than the core flux density of Lh09c1 returned {\it getsf}, 0.048~Jy, which only accounts for the compact part. This suggests that the 0.87~mm emission from Lh09c1, as well as its surroundings within 0\farcs{25} (0.05~pc), is dominated by free-free emission. 

On the larger clump scale, we adopt the MAGOS 0.1~pc resolution data. Within a radius of 0\farcs{8} (0.2~pc), the 0.87~mm flux density is 0.322~Jy, while the H37$\gamma$ flux density measured from the 0.1~pc resolution data at the same aperture is 1.89~Jy\kms, corresponding to a free-free continuum contribution of 0.170~Jy. Therefore, the dust contribution on the clump scale is approximately 0.152~Jy.



\bibliography{refer}



\end{document}